\documentclass[10pt,orivec]{llncs}
\usepackage{amssymb}
\usepackage{amsmath}
\usepackage{color}
\usepackage{array}
\usepackage{longtable}
\usepackage{booktabs}
\usepackage{verbatim}
\usepackage{cite}
\usepackage[noend]{algpseudocode}
\usepackage{algorithm,algorithmicx}
\usepackage{subfigure}
\usepackage{multirow}
\usepackage{hyperref}
\usepackage{quantikz}
\usepackage{rotating}
\hypersetup{
    colorlinks=true,
    linkcolor=blue,
    filecolor=blue,      
    urlcolor=blue,
    citecolor=blue,
}

\newcommand{\setnocond}[1]{\{#1\}}
\newcommand{\setcond}[2]{\{\,#1 : #2\,\}}

\newcommand{\e}{\ensuremath{\mathcal{E}}}
\newcommand{\h}{\ensuremath{\mathcal{H}}}
\newcommand{\tr}{\ensuremath{\mathrm{tr}}}

\newcommand{\coloneqq}{:=}
\newcommand{\Coloneqq}{::=}

\renewcommand{\bra}[1]{\langle #1 |}
\renewcommand{\ket}[1]{| #1 \rangle}
\renewcommand{\braket}[2]{\langle #1 | #2 \rangle}
\newcommand{\ketbra}[2]{| #1 \rangle \langle #2 |}

\renewcommand{\orcidID}[1]{\smash{\href{http://orcid.org/#1}{\protect\raisebox{-1.25pt}{\protect\includegraphics{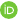}}}}}

\title{Model Checking Matrix Product States against Linear Chain Logic}

\author{Ming Xu\inst{1}\orcidID{0000-0002-9906-5677} \and
Yihao Chen\inst{1}\orcidID{0009-0006-2235-1276} \and
Ji Guan\inst{2}\thanks{Corresponding author.}\orcidID{0000-0002-3490-0029}
}

\institute{School of Cryptology, East China Normal University, Shanghai 200062, China \and
Key Laboratory of System Software (Chinese Academy of Sciences),
Institute of Software, Chinese Academy of Sciences, Beijing 100190, China \\
\email{mxu@cs.ecnu.edu.cn, yihaochen0513@gmail.com, guanj@ios.ac.cn}
}

\begin{document}
\maketitle
\pagestyle{plain}

\begin{abstract}
    Matrix product states (MPS) are a standard tensor-network representation
    for ground states of one-dimensional quantum many-body systems,
    and they underpin widely used simulation tools such as DMRG.
    However, while quantum model checking has been developed mainly for quantum programs and communication protocols
    (with properties expressed along a time axis),
    there is still no comparable framework for systematically verifying \emph{spatial}
    and \emph{size-dependent} properties of physical many-body states,
    where the key parameter is the system size.
    This paper takes a step toward bridging the gap.
    We propose \emph{Linear Chain Logic} (LCL),
    a spatial logic designed to specify physically meaningful properties of periodic MPS families
    as the system size grows,
    such as nontriviality on rings and large-size asymptotic patterns.
    Our approach builds on a simple but powerful connection:
    every periodic MPS naturally induces a completely positive map (a quantum operation) on its virtual space,
    so many quantitative features of the MPS can be analysed through the repeated application of the operation.
    Using this perspective,
    we derive an effective procedure to compute the inner products of an MPS at a given size
    and to support richer LCL specifications,
    without relying on brute-force state expansion.
    We then develop approximate model-checking algorithms that combine sound bounding
    with asymptotic structural analysis,
    enabling scalable reasoning about large system sizes.
    Experiments on representative MPS families illustrate that
    our method can automatically verify nontriviality and detect asymptotic spatial regimes
    in a way that complements traditional numerical techniques.
\end{abstract}

\section{Introduction}
Model checking is a cornerstone of formal verification:
given a \emph{system model} and a \emph{specification},
it decides whether the former satisfies the latter~\cite{CGP99,baier2008principles}.
Driven by quantum software stacks and quantum networks,
model checking has been extended to quantum settings and become an effective tool
for validating quantum programs and communication protocols~\cite{ying2019model,ying2021model}.
As in the classical case, success hinges on two ingredients:
i)~a precise model of system evolution,
and ii)~a property language with an executable checking procedure.

\paragraph{Where quantum model checking has been most successful.}
Most existing frameworks target \emph{temporal} verification tasks
that arise naturally in computation and communication:
program steps, rounds, and time slots.
A representative model is the \emph{quantum Markov chain} (QMC),
which generalises classical Markov chains
by replacing stochastic transitions with completely positive maps (quantum operations)~\cite{wolf2012quantum}.
This paradigm underlies verification across quantum control~\cite{Lidar2012},
quantum information theory~\cite{guan2018structure},
quantum programming~\cite{ying2024foundations},
and quantum communication protocols~\cite{ying2019model,ying2021model}.

\paragraph{A growing verification frontier: state families indexed by size.}
Many important quantum systems are not naturally described as a time-evolving protocol,
but as a \emph{family of states} parameterised by a size parameter
(e.\,g.\@, the number of sites $N$ in a one-dimensional (1D) chain)~\cite{zeng2015quantum}.
The viewpoint is central in quantum simulation and device-level validation,
where the goal is to \emph{prepare and certify} target states and their properties as the system scales.
From a verification perspective, the core challenge is therefore \emph{spatial} and \emph{asymptotic}:
properties may stabilize only beyond a threshold, recur periodically, or change regime as $N$ grows.
It calls for a model-checking perspective that reasons over the index $N$, rather than over time.
Recent work at the interface of automata/regular languages and quantum state representations further suggests that
formal-methods structure can fruitfully organise many-body state classes and their properties~\cite{florido2024regular}.

\paragraph{MPS as a scalable system model.}
For 1D local quantum systems,
the \emph{matrix product state} (MPS) is a standard compact representation,
which underpins numerical methods such as DMRG~\cite{schollwock2011dmrg}.
From a verification perspective,
the key point is that many physically relevant 1D ground states admit such compact descriptions:
for gapped local Hamiltonians, entanglement satisfies an entanglement area law~\cite{hastings2007area},
which implies approximability by MPS with a bond dimension
that grows only with the desired accuracy (and model parameters),
rather than exponentially with the chain length~\cite{aharonov2011area}.
Thus, MPS provides a natural \emph{system model} for scalable,
size-indexed reasoning about state families $\{\ket{\psi_N}\}_{N\ge 1}$
generated by a fixed set of matrices $\{A_k\}_{k=1}^d$ (to be formalised in Section~\ref{sec:pre}):
\begin{equation}
    \ket{\psi_N}
    =\sum_{k_1,\ldots,k_N=1}^{d}
    \tr(A_{k_1}\cdots A_{k_N})\;
    \ket{k_1}\otimes\cdots\otimes\ket{k_N}.
\end{equation}

\paragraph{The gap: model checking for spatial, quantitative properties of MPS families.}
Despite substantial progress on temporal logics and QMC-style models,
systematic verification of \emph{spatial} properties of MPS families remains largely unexplored.
Yet these properties are exactly what one needs to study and validate quantum many-body behavior:
Is the periodic MPS nontrivial for all sufficiently large size parameter $N$?
Do key observables eventually lie in a prescribed range?
Do behaviors become ultimately periodic or converge to a limiting regime?
Answering such questions requires: i)~a logic whose semantics ranges over the size parameter $N$,
and ii)~algorithms that exploit the algebraic structure of the MPS tensors.

\paragraph{Our approach.} To fill this gap,
we propose a spatial logic, \emph{Linear Chain Logic} (LCL),
and approximate model-checking algorithms for periodic MPS.
The key technical bridge is that
a fixed MPS tensor family $\{A_k\}$ induces a completely positive (CP) map
$\mathcal{E}(X)=\sum_k A_k X A_k^\dagger$ on the bond space (the space of $A_k$),
together with the (Liouville) matrix representation $M_{\mathcal{E}}=\sum_k A_k^*\otimes A_k$.
The transfer-operator view turns many MPS quantities into trace expressions of matrix powers.
Particularly, $\|\psi_N\|^2=\tr(M_{\mathcal{E}}^{\,N})$,
so fundamental questions are reduced to reasoning about the scalar sequence
$\{\tr(M_{\mathcal{E}}^{\,N})\}_{N\ge 1}$.
Furthermore,
structural results for irreducible CP maps constrain the peripheral spectrum to be cyclic (roots of unity),
yielding eventual periodic regimes that our algorithms can exploit
without requiring an explicit full-spectrum computation.

\paragraph{Logic and algorithms.}
LCL is LTL-like in syntax, but interpreted over the size index $N$.
Value expressions $\Gamma(N)$ are extracted from the MPS contractions,
interval predicates $\in \mathcal{I}$ give the ranges,
atomic formulas combine them as $\Gamma(N) \in \mathcal{I}$,
and general formulas admit Boolean connectives and spatial operators
such as ``eventually'' and ``infinitely often''
along $N$.
Our approximate model-checking pipeline:
i)~computes sound bounds for $\Gamma(N)$ using the transfer-operator structure,
ii)~derives semilinear over-/under-approximations of the set of sizes satisfying each atomic formula,
and iii)~propagates these approximations bottom-up to decide the input formula.
In sum, this paper makes three main contributions:
\begin{enumerate}
\item \textbf{Specification} --- We introduce the \emph{linear chain logic} (LCL)
to express size-indexed, spatial properties of periodic MPS families
in many-body systems.
\item \textbf{Technique} --- We develop the transfer-operator-based model checking,
leveraging CP-map view, irreducible decomposition,
and peripheral cyclicity to compute semilinear over-/under-approximations efficiently.
\item \textbf{Validation} --- We evaluate the checker on representative MPS families and quantitative properties,
demonstrating scalable certification of nontriviality and detection of asymptotic spatial regimes.
\end{enumerate}

\subsection{Related work}\label{sec:related}

\paragraph{(1) Chain logic vs. temporal logics:}
Classical model checking is tightly coupled with temporal logics,
where the system is observed along a time axis.
Linear-time logics such as LTL 
specify properties over single executions,
while branching-time logics such as CTL/CTL$^\ast$ quantify over computation trees and enable richer path quantification
(e.\,g.\@, reachability)~\cite{clarke1986automatic,baier2008principles}.
The logics and their model-checking pipelines are now standard for software/hardware verification.
Our setting differs in what the ``axis'' ranges over:
rather than time steps of a program,
we consider the spatial index of a family of periodic MPS $\{\ket{\psi_N}\}_{N\ge1}$,
and the target properties are quantitative and asymptotic in the chain length $N$.
It motivates our LCL to allow LTL-like connectives/operators,
but interpreted over the size parameter $N$ and grounded in MPS contractions
(e.\,g.\@, norms/correlators) as value expressions.
Thus, LCL fills a gap not covered by existing temporal logics:
it provides a systematic specification language for \emph{spatial} size-indexed properties of
tensor-network state families, together with a dedicated checking pipeline.

\paragraph{(2) From quantum computing and communication models to physical many-body systems.}
Most quantum model-checking work has been developed for computational and communication models,
where evolution is naturally temporal and often formalised
via quantum automata and quantum Markov chains (QMC)
driven by quantum operations and channels~\cite{feng2013model,ying2019model,ying2018model}.
In parallel,
recent years have seen a growing interplay between formal-language or automata viewpoints and quantum physics,
including explicit ``language-theoretic'' representations of quantum many-body states
(e.\,g.\@ regular language quantum states)~\cite{florido2024regular}.
These developments highlight an important frontier:
bringing verification ideas beyond program-like temporal evolution to \emph{many-body state families}
where the central object is a ground state ansatz
and the key questions are quantitative properties across system sizes
(nontriviality on rings, asymptotic regimes, thresholds, eventual periodicity).
Our paper contributes to this broadening by
i)~taking \emph{periodic MPS} as the system model for 1D many-body physics,
and ii)~exploiting the ``highway'' connection from MPS tensors to CP maps (transfer operators),
so that MPS verification reduces to reasoning about algebraic objects
(CP maps and channels) familiar from QMC theory,
while the specification layer is redesigned to be spatial rather than temporal.

\paragraph{(3) Traditional approaches for analysing or certifying MPS.}
MPS are the backbone of 1D many-body numerics and theory:
DMRG and MPS methods and tensor-network tooling provide efficient computation of energies and local observables,
and the transfer-matrix formalism enables scalable evaluation of expectation values
and correlations~\cite{schollwock2011dmrg,orus2014practical,Perez-Garcia2007,fannes1992finitely}.
Canonical forms, injectivity and parent-Hamiltonian techniques,
and spectral properties of transfer operators underpin much of the existing ``verification'' practiced in physics:
one typically checks consistency by computing energies, correlation sizes, or matching known invariants,
often for fixed sizes or in the thermodynamic limit under additional assumptions.
The gap is that these toolchains do not provide a \emph{logic-based},
property-driven model-checking interface for \emph{size-indexed quantitative specifications}
over the entire family $\{\ket{\psi_N}\}_{N\ge1}$,
nor do they directly target decision problems
like eventual nonzeroness or dichotomy under periodic boundary conditions.
Our contribution is complementary:
we build a verification layer on top of the transfer-operator and CP-map structure,
developing approximate model checking that returns semilinear over/under-approximations of satisfying sizes
and systematically composes such results according to an input spatial formula (LCL),
enabling scalable certification of nontriviality and asymptotic spatial regimes beyond ad-hoc observable computations.

\section{Preliminaries}\label{sec:pre}
This section recalls quantum states and Hamiltonian-governed evolution,
and then explains why ground states of 1D local many-body Hamiltonians are commonly represented
by matrix product states (MPS).
We finally define the \emph{periodic} MPS studied in this paper and state the nonzeroness verification problem
that motivates our logic-based analysis.
For more background, see~\cite{nielsen2010quantum,cirac2021matrix}. 

\subsection{Quantum states}\label{subsec:pre_states}

\paragraph{State space $\h$.}
A single quantum particle (or local degree of freedom) is described
by a finite-dimensional complex Hilbert space $\h\cong\mathbb{C}^d$, where $d=\dim(\h)$.
The notion $\ket{\psi}$ denotes a (column) vector in $\h$,
which is not necessarily unit considered in this paper.
Only pure (quantum) states are referred to unit vectors $\ket{\psi}\in\h$.
Fix an orthonormal computational basis $\{\ket{1},\ldots,\ket{d}\}$ of $\h$.
Then any pure state $\ket{\psi}$ admits the expansion
\begin{equation}
    \ket{\psi}=\sum_{k=1}^{d} a_k \ket{k}
    \quad \text{with} \quad
    \sum_{k=1}^{d} |a_k|^2=1.
\end{equation}
We use Dirac notation (standard in quantum computing) throughout:
\begin{enumerate}
    \item $\bra{\psi}\coloneqq \ket{\psi}^\dagger$ (conjugate transpose);
    \item $\braket{\psi_1}{\psi_2}\coloneqq \bra{\psi_1}\ket{\psi_2}$ (inner product);
    \item $\ket{\psi_1,\psi_2}\coloneqq \ket{\psi_1}\otimes\ket{\psi_2}$ (tensor product).
\end{enumerate}
For a linear operator (matrix) $X$, we write $X^\dagger$ for the conjugate transpose
and $\tr(X)$ for the trace (sum of diagonal entries).
For example, in a 1-qubit circuit model~\cite{nielsen2010quantum},
the state space is 2-dimensional,
i.\,e.\@, $\h\cong\mathbb{C}^2$ with computational basis $\{\ket{1},\ket{2}\}$,
where $\ket{1}=(1,0)^{\mathsf T}$ and $\ket{2}=(0,1)^{\mathsf T}$.
A 1-qubit pure state has the form
$\ket{\psi}=\alpha\ket{1}+\beta\ket{2}$ with $\alpha,\beta\in\mathbb{C}$,
satisfying $|\alpha|^2+|\beta|^2=1$.

\paragraph{Tensor-product state space.}
Consider a collection of $N$ quantum particles,
each with local Hilbert space $\h\cong\mathbb{C}^d$.
The global state space is the tensor power $\h_{\mathrm{all}}=\h^{\otimes N}$,
whose dimension is $d^N$ and thus grows exponentially with $N$.
A pure state on $\h_{\mathrm{all}}$ can be expanded in the computational basis
\begin{equation}
    \Big\{\ket{i_1}\otimes\ket{i_2}\otimes\cdots\otimes\ket{i_N}\ \Big|\ i_j\in\{1,\ldots,d\}\Big\},
\end{equation}
highlighting the intrinsic challenge of representing
and reasoning about large many-body (multiple-particle) quantum states.

\subsection{Hamiltonian evolution and ground states}\label{subsec:pre_ham}

\paragraph{Hamiltonian evolution.}
The evolution of an isolated quantum system is governed by a \emph{Hamiltonian} $H=H^\dagger$ acting on $\h$.
The time evolution of a pure state follows Schr\"odinger's differential equation
$i\,\frac{d}{dt}\ket{\psi(t)}=H\ket{\psi(t)}$,
whose solution is unitary:
$\ket{\psi(t)}=U(t)\ket{\psi(0)}$ with $U(t)=e^{-itH}$.
Some popular 1-qubit (2-dimensional) Hamiltonian (matrices) are the Pauli matrices
\[
    I=\begin{bmatrix}
        1 & 0 \\
        0 & 1
    \end{bmatrix},\qquad X=\begin{bmatrix}
        0 & 1 \\
        1 & 0
    \end{bmatrix},\qquad Y=\begin{bmatrix}
        0 & -i \\
        i & 0
    \end{bmatrix},\qquad Z=\begin{bmatrix}
        1 & 0 \\
        0 & -1
    \end{bmatrix},
\]
and the Hadamard matrix $\mathbf{H}=(X+Z)/\sqrt{2}$.
Additionally, the spin-$\tfrac12$ matrices are $S^x=\tfrac{1}{2} X$,
$S^y=\tfrac{1}{2} Y$ and $S^z=\tfrac{1}{2} Z$.
They will be used to composite complicated Hamiltonian in our experiments.

\paragraph{Local Hamiltonians and ground states (1D chains).}
A one-dimensional (1D) quantum many-body system consists of many (typically $N\gg 1$)
interacting quantum degrees of freedom (e.\,g.\@, spins, atoms, or electrons arranged on a line),
whose joint state space is $\h^{\otimes N}$.
We focus on finite-range, translationally invariant 1D Hamiltonians of the form
$H_N=\sum_{i=1}^{N} h_{i,n}$,
where each local term $h_{i,n}$ is centered at site $i$ and acts nontrivially only on the $(n-1)$ nearest sites
(interaction range $n$). 
Let $E_0(N)=\min\mathrm{spec}(H_N)$ be the ground energy.
A normalised pure state $\ket{\psi}$ is a \emph{ground state} if $H_N\ket{\psi}=E_0(N)\ket{\psi}$.
Ground states play a central role in many-body physics: at low temperature,
equilibrium behavior is dominated by the ground space and the lowest-energy excitations,
and macroscopic order and correlations are largely reflected in ground-state structure.

\subsection{Matrix product state representation}\label{subsec:pre_mps}

\paragraph{Why MPS for 1D ground states.}
Although $\h^{\otimes N}$ has exponential dimension,
1D ground states of local Hamiltonians often admit compact descriptions due to limited entanglement.
In particular, for broad classes of gapped 1D local Hamiltonians,
ground states satisfy an entanglement area law~\cite{hastings2007area},
which implies they can be approximated to arbitrary accuracy by MPS with sufficiently large bond dimension $D$,
with quantitative trade-offs between the bond dimension and the target error~\cite{aharonov2011area}.
Importantly, the required $D$ is not determined solely by the interaction range:
it is governed by the amount of entanglement across bipartitions,
where for an exact MPS one has the entropy bound $S\le \log D$.

\begin{definition}[Periodic matrix product state]\label{def:periodic-mps-pre}
Let $\{\ket{1},\ldots,\ket{d}\}$ be an orthonormal basis of $\h\cong\mathbb{C}^d$.
Given a set of matrices $\{A_k\}_{k=1}^{d} \subset \mathbb{C}^{D \times D}$,
the \emph{periodic matrix product state} (MPS) on $N$ sites is
\begin{equation}\label{eq:periodic-mps-pre}
    \ket{\psi_N} \coloneqq
    \sum_{k_1,\ldots,k_N=1}^{d} \tr(A_{k_1}\cdots A_{k_N})\, \ket{k_1}\otimes\cdots\otimes\ket{k_N}.
\end{equation}
Here, $\{A_k\}_{k=1}^{d}$ are called the tensors of the MPS,
and the virtual space that the matrices $A_k$ act on is the bond space,
whose dimension $D$ is independent of $N$.
\end{definition}
With the tensors, we can define a family of MPS $\{\ket{\psi_N}\}_{N\geq 1}$,
which is determined by $d \cdot D^2$ parameters. 

The trace in~\eqref{eq:periodic-mps-pre}
identifies the two boundary virtual indices and closes the tensor network into a ring,
i.\,e.\@, it imposes \emph{periodic boundary conditions}.
Without that trace, it is led to \emph{the boundary conditions}
$\sum_{k_1,\ldots,k_N} A_{k_1}\cdots A_{k_N}\, \ket{k_1}\otimes\cdots\otimes\ket{k_N}$.
Throughout this paper we focus on \emph{periodic} matrix product states;
for brevity, we will refer to them simply as matrix product states (MPS).

\begin{figure}[ht]
  \centering
  \includegraphics[width=0.6\linewidth]{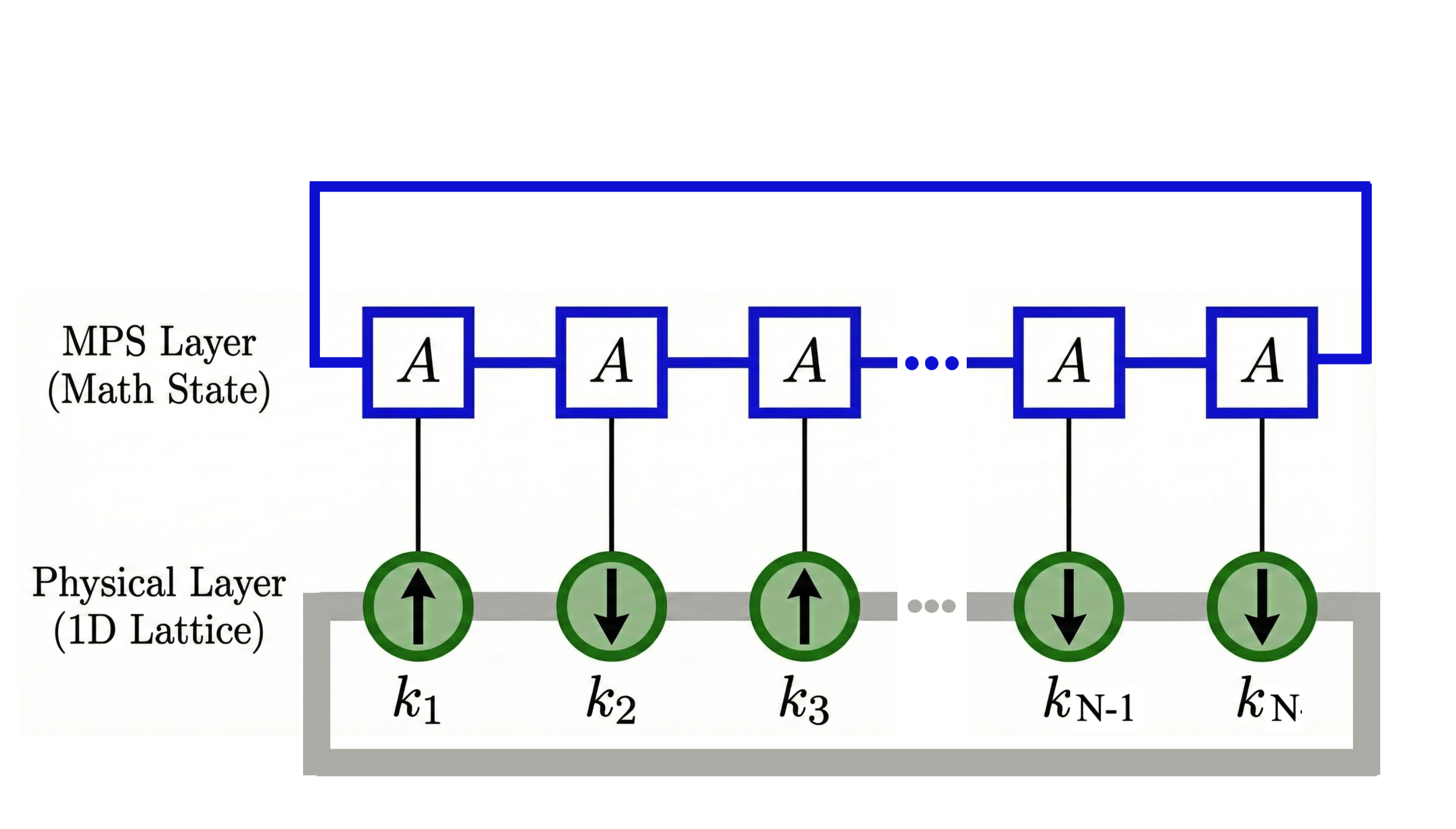}
  \caption{Schematic overview of a 1D quantum many-body system and its MPS.}
  \label{fig:1d-mps}
\end{figure}

\paragraph{Ring-contraction view.}
Equation~\eqref{eq:periodic-mps-pre} can be seen as a closed tensor-network contraction on an $N$-site ring,
as shown in Fig.~\ref{fig:1d-mps}.
Accordingly, $\ket{\psi_N}\neq 0$ means that this closed contraction does not vanish identically,
i.\,e.\@, the local tensors $\{A_k\}$ define a nontrivial many-body wavefunction on a ring of length $N$.

\begin{problem}[MPS Nonzeroness Dichotomy]\label{pro:many_body}
Given a set of matrices $\{A_k\}_{k=1}^{d}\subset\mathbb{C}^{D\times D}$
defining $\ket{\psi_N}$ in~\eqref{eq:periodic-mps-pre}
and an integer $J\ge 1$, decide:
\begin{itemize}
\item (Always-nonzero after $J$) Does $\ket{\psi_N}\neq 0$ hold for all $N\ge J$?
\item (Ultimately nonzero) Does there exist $N_0\ge J$ such that $\ket{\psi_N}\neq 0$ holds for all $N\ge N_0$?
\end{itemize}
\end{problem}

\paragraph{Physical meaning.}
Problem~\ref{pro:many_body} asks
whether a fixed local tensor $\{A_k\}$ yields a valid periodic many-body state on all sufficiently large rings,
or whether exact cancellations can force the periodic contraction to vanish for infinitely many system sizes.
This ``eventual nontriviality'' property is a natural target for \emph{spatial} verification over the size axis $N$,
and will be expressed and checked using Linear Chain Logic proposed in Section~\ref{sect:logic}.

\begin{example}\label{ex1}
We consider a 1D translation-invariant MPS with local dimension $d=3$ and bond dimension $D=4$
specified by the matrices $A_1,A_2,A_3 \in \mathbb{C}^{4 \times 4}$:
\[
A_1=
\begin{bmatrix}
0 & 0 & \alpha_1 & \alpha_2 \\[4pt]
0 & 0 & \alpha_2 & \alpha_1 \\[4pt]
\alpha_1 & \alpha_2 & 0 & 0 \\[4pt]
\alpha_2 & \alpha_1 & 0 & 0
\end{bmatrix},
\qquad
A_2=
\begin{bmatrix}
0 & 0 & \alpha_3 & \alpha_4 \\[4pt]
0 & 0 & \alpha_4 & \alpha_3 \\[4pt]
\alpha_3 & \alpha_4 & 0 & 0 \\[4pt]
\alpha_4 & \alpha_3 & 0 & 0
\end{bmatrix},
\qquad
A_3=
\begin{bmatrix}
\alpha_5 & \alpha_6  & 0 & 0 \\[4pt]
\alpha_6 & \alpha_5  & 0 & 0 \\[4pt]
0 & 0 & \alpha_6 & \alpha_5  \\[4pt]
0 & 0 & \alpha_5 & \alpha_6
\end{bmatrix},
\]
where $\alpha_1 = \frac{\sqrt{3}}{6}+\frac{\sqrt{2}}{4}$,
$\alpha_2  = -\frac{\sqrt{3}}{6}+\frac{\sqrt{2}}{4}$,
$\alpha_3 = \frac{\sqrt{3}\varpi}{6}+\frac{\sqrt{2}}{4}$,
$\alpha_4  = -\frac{\sqrt{3}\varpi}{6}+\frac{\sqrt{2}}{4}$,
$\alpha_5 = \frac{\sqrt{3}}{6}$
and $\alpha_6  = -\frac{\sqrt{3}}{6}$ ($\varpi$ denotes the 8th root of unity).
Then we get the famliy of MPS $\{\ket{\psi_N}\}_{N \ge 1}$,
in which $\ket{\psi_1} = 0$,
$\ket{\psi_2} = \tfrac{5}{3}\ket{1,1}-\tfrac{1+2i}{3}\ket{2,2} +\tfrac{2}{3}\ket{3,3}
+ (1+\tfrac{2}{3}\varpi)(\ket{1,2}+\ket{2,1})$,
and so on.
The Dichotomy problem concerns
whether the MPS sequence $\ket{\psi_N}$ is (ultimately) nonzero after a threshold $J$. \qed
\end{example}

\subsection{Connection to quantum channel}\label{subsec:channel}
A key bridge between MPS and our verification procedure is the \emph{transition super-operator}
induced by the local MPS tensors.
Given an MPS specified by matrices $\{A_k\}_{k=1}^{d}\subset\mathbb{C}^{D\times D}$,
we define a linear map $\e:\mathbb{C}^{D\times D}\to \mathbb{C}^{D\times D}$
as in~\cite{de2017irreducible}:
\begin{equation}\label{eq:mps-channel}
    \e(X) \ \coloneqq \ \sum_{k=1}^{d} A_k \,X\, A_k^\dagger .
\end{equation}
By construction, $\e$ is completely positive (CP),
since it admits a Kraus form as in the Kraus representation theorem~\cite{choi1975completely}.
The matrices $\{A_k\}$ are called the Kraus operators of $\e$. 
Moreover, if the tensors satisfy the \emph{trace-preserving} condition $\sum_{k=1}^{d} A_k^\dagger A_k = I_D$,
$\e$ is a \emph{quantum channel} (CPTP map).
In this case, $\e^N$ corresponds to $N$ repeated steps of the same process.
Equivalently, $(\mathbb{C}^D,\e)$ can be viewed as
a \emph{quantum Markov chain}~\cite{feng2013model} with transition super-operator $\e$.
Finally, the periodic contraction of the MPS admits a convenient expression through this transfer map:
many scalar observables relevant to our logic (e.\,g.\@, norms and correlation functions)
reduce to quantities of the form $\tr(M^N)$,
where $M$ is the matrix representation of $\e$,
which is detailed in Section~\ref{sec:nonzeroness}.
The reduction enables spatial verification of MPS by analysing the iterates of $\e$.

\section{Inner Product of MPS}\label{sec:nonzeroness}

In this section, as a first step toward solving the dichotomy problem (Problem~\ref{pro:many_body}),
we study the inner product of MPS $\ket{\psi_N}$ for a given $N$.
It also serves as a fundamental subroutine for our model-checking procedures in Section~\ref{sect:algo},
where we lift single-size reasoning to \emph{asymptotic} guarantee, including the dichotomy problem and
other physically meaningful properties expressible in our logic.

Firstly, we reduce the nonzeroness of a periodic MPS to a scalar quantity derived directly from its local tensors.
Since $\ket{\psi_N}=0$ if and only if its squared norm $\|\psi_N\|^2=\braket{\psi_N}{\psi_N}$ is zero,
our nonzeroness questions boil down to reasoning about a scalar sequence computed from the MPS data.

Let $\mathcal{E}$ be the CP map induced by the MPS tensors $\{A_k\}_{k=1}^d$.
Then, the Liouville matrix representation~\cite{ying2018model} is
\begin{equation}\label{eq:def-superop}
    M_{\mathcal{E}}\ \coloneqq \ \sum_{k=1}^{d} A_k^{*}\otimes A_k
    \quad \text{for} \quad
    \mathcal{E}(X) = \sum_{k=1}^{d} A_k\,X\,A_k^\dagger,
\end{equation}
where $A^*$ denotes entry-wise complex conjugation.
A direct tensor-network contraction (or a trace manipulation) yields:%
{
\begin{align*}
\braket{\psi_N}{\psi_N}
&= \sum_{k_{1},\dotsc,k_{N}=1}^{d}
   \tr(A_{k_{N}}^\dagger \cdots A_{k_{1}}^\dagger)\;\tr(A_{k_{1}} \cdots A_{k_{N}}) \\
&= \sum_{k_{1},\dotsc,k_{N}=1}^{d}
   \tr(A_{k_{N}}^{*} \cdots A_{k_{1}}^{*})\;\tr(A_{k_{1}} \cdots A_{k_{N}}) \displaybreak[0] \\
&= \sum_{k_{1},\dotsc,k_{N}=1}^{d}
   \tr\Big((A_{k_{N}}^{*} \cdots A_{k_{1}}^{*})\otimes(A_{k_{1}}\cdots A_{k_{N}})\Big) \displaybreak[0] \\
&= \sum_{k_{1},\dotsc,k_{N}=1}^{d}
   \tr\Big((A_{k_1}^{*}\!\otimes\! A_{k_1})\cdots(A_{k_N}^{*}\!\otimes\! A_{k_N})\Big) \displaybreak[0] \\
&= \tr\Big(\big[\sum_{k=1}^{d}A_{k}^{*}\!\otimes\! A_{k}\big]^N\Big)
 = \tr\big(M_{\mathcal{E}}^{\,N}\big).
\end{align*}
}%
Therefore, the squared norms of the MPS family $\{\ket{\psi_N}\}_{N\ge 1}$ form the scalar sequence
$\{\tr(M_{\mathcal{E}}^{\,N})\}_{N\ge 1}$.
Equivalently, they are determined by the discrete-time quantum evolution on the bond space
$(\mathcal{H}_{\mathrm{bond}},\mathcal{E})$ with $\mathcal{H}_{\mathrm{bond}}\cong\mathbb{C}^D$.
This reduction allows us to explore $\|\psi_N\|$ symbolically via linear-algebraic and spectral properties of
$M_{\mathcal{E}}$ (or $\mathcal{E}$), and will be the basis of our model-checking algorithms.

To understand the long-run behavior of $\tr(M_{\mathcal{E}}^{\,N})$ and hence the asymptotic validity
and periodic phenomena of $\ket{\psi_N}$,
we analyse the spectral structure of $M_{\mathcal{E}}$
(or equivalently of $\mathcal{E}$ under vectorization).

\paragraph{Peripheral eigenvalues.}
By standard linear algebra~\cite[Sect.~7.3]{HK71}, we have that
every (square) matrix $X$ is unitarily similar to a Jordan canonical form
with all eigenvalues (counted with algebraic multiplicity) of $X$ as diagonal entries,
and thereby the trace of $X$ amounts to the sum of all eigenvalues.
Writing the eigenvalues of $M_{\mathcal{E}}$
as $\lambda_1,\ldots,\lambda_{D^2}$ yields
\begin{equation}\label{eq:trace-spectral-sum}
    \tr(M_{\mathcal{E}}^{\,N})=\sum_{i=1}^{D^2}\lambda_i^{N}.
\end{equation}
Eigenvalues of maximal modulus $\rho(M_{\mathcal{E}})$ are called \emph{peripheral}.
When $N$ is large, the terms $\lambda^N$ with $|\lambda|=\rho(M_{\mathcal{E}})$ dominate the sum,
and their arguments determine whether $\tr(M_{\mathcal{E}}^{\,N})$ exhibits oscillations or periodic patterns.
However, clean structural characterizations of the peripheral spectrum are available primarily for
\emph{irreducible} CP maps. We therefore reduce $\mathcal{E}$ to an irreducible form while preserving
$\tr(M_{\mathcal{E}}^{\,N})$.

\paragraph{Invariant subspaces and block reduction.}
Let $\mathcal{H}\cong\mathbb{C}^D$ be the bond space.
A subspace $\mathcal{H}_1\subseteq \mathcal{H}$ is \emph{invariant} for $\mathcal{E}$
if $A_k \mathcal{H}_1\subseteq \mathcal{H}_1$ for all $k$.
Then, we have:
\begin{lemma}[Refinement]\label{lem:refine}
    Let $\h_1$ be an invariant subspace for $\e$,
    and $\h_2$ the orthogonal complement.
    Then the MPS defined by the Kraus operators of $\e$ can be decomposed
    into the two MPS defined by the Kraus operators restricted to $\h_1$ and to $\h_2$, respectively.
\end{lemma}

\paragraph{Irreducible decomposition.}
By iterating the refinement with Lemma~\ref{lem:refine},
we obtain an orthogonal decomposition
$\mathcal{H}=\bigoplus_{m} \mathcal{H}_m$
such that each $\mathcal{H}_m$ is invariant for $\mathcal{E}$ and the restriction
$\mathcal{E}_m \coloneqq \mathcal{E}\!\mid_{\mathcal{H}_m}$ is \emph{irreducible},
i.\,e.\@,
there is no nontrivial projector $P$ on $\mathcal{H}_m$ with
$\mathcal{E}_m(\mathcal{L}(P\mathcal{H}_m))\subseteq \mathcal{L}(P\mathcal{H}_m)$,
where $\mathcal{L}(\mathcal{H}_m)$ means the set of linear operators on $\mathcal{H}_m$.
Let $P_m$ denote the projector onto $\mathcal{H}_m$, and write $B_{m,k} \coloneqq P_m A_k P_m$.
Define the associated MPS on each block by
\begin{equation}
    \ket{\phi_{N,m}}
    =\sum_{k_1,\ldots,k_N=1}^{d}
    \tr(B_{m,k_1}\cdots B_{m,k_N})\,\ket{k_1}\cdots\ket{k_N}.
\end{equation}
Then the MPS decomposes additively:
\begin{equation}\label{eq:equiv}
    \ket{\psi_N}=\sum_m \ket{\phi_{N,m}} \qquad (N\ge 1).
\end{equation}
We call $\{(\mathcal{H}_m,\mathcal{E}_m)\}_m$
an \emph{irreducible form} of $\mathcal{E}$~\cite{Cirac2017,de2017irreducible}.
We provide the construction steps for analysing the complexity of our method afterwards. 
This reduction isolates the components that govern the asymptotic behavior and enables a sharp
characterization of peripheral periodicity.

\begin{example}\label{ex3}
We reconsider the bond space $\mathcal{H}=\mathrm{span}\{\ket{1},\ket{2},\ket{3},\ket{4}\}$
and the CP map $\mathcal{E}(X)=\sum_{k=1}^3 A_k \,X\, A_k^\dagger$ from Example~\ref{ex1}.
One can verify that the subspaces
\[
    \mathcal{H}_1=\mathrm{span}\Big\{\tfrac{\ket{1}-\ket{2}}{\sqrt2},\,\tfrac{\ket{3}-\ket{4}}{\sqrt2}\Big\},
    \qquad
    \mathcal{H}_2=\mathrm{span}\Big\{\tfrac{\ket{1}+\ket{2}}{\sqrt2},\,\tfrac{\ket{3}+\ket{4}}{\sqrt2}\Big\},
\]
are invariant under all $A_k$ and are irreducible.
Let $U$ be the unitary change-of-basis matrix whose columns are the above orthonormal vectors.
Then $U^\dagger A_k U$ is block-diagonal with blocks $B_{1,k}$ and $B_{2,k}$, and for any word $(k_1,\ldots,k_N)$,
\[
\begin{aligned}
    \tr(A_{k_1}\cdots A_{k_N})
    & =\tr\big((U^\dagger A_{k_1}U)\cdots(U^\dagger A_{k_N}U)\big) \\
    & =\tr(B_{1,k_1}\cdots B_{1,k_N})+\tr(B_{2,k_1}\cdots B_{2,k_N}).
\end{aligned}
\]
This entails~\eqref{eq:equiv} with two components $\ket{\phi_{N,1}}$ and $\ket{\phi_{N,2}}$. \qed
\end{example}

\paragraph{Peripheral periodicity for irreducible CP maps.}
A key structural fact --- often phrased for CP maps with unit spectral radius ---
is that \emph{irreducibility enforces cyclicity} of the peripheral spectrum.

\begin{lemma}[\cite{de2017irreducible, wolf2012quantum}]\label{lem:cyclic}
Let $\mathcal{E}$ be an irreducible CP map on $\mathcal{L}(\mathcal{H})$
with spectral radius $\rho(\mathcal{E})=1$ and $\dim(\mathcal{H})=D$.
Then:
\begin{itemize}
\item $1$ is a simple peripheral eigenvalue of $\mathcal{E}$; and
\item if $\mathcal{E}$ has other peripheral eigenvalues, then there exists an integer $p\le D^2$
such that the peripheral eigenvalues are exactly the $p$th roots of unity.
\end{itemize}
\end{lemma}
Lemma~\ref{lem:cyclic} implies that, for each irreducible component, the asymptotically dominant part of
$\tr(M_{\mathcal{E}}^{\,N})$ is governed by roots of unity.
Consequently, once subdominant eigenmodes (with modulus strictly smaller than~$1$) become negligible,
$\tr(M_{\mathcal{E}}^{\,N})$ exhibits \emph{periodic/oscillatory asymptotic patterns} with period dividing~$p$.
We will exploit this peripheral periodicity in our model-checking procedures, while avoiding explicit computation
of the full spectrum (which can be much expensive in general~\cite[Algorithm~10.4]{BPR06}).

\begin{example}
For the irreducible restrictions in Example~\ref{ex3}, the spectra are
$\mathrm{spec}(\mathcal{E}_1) \linebreak[0]=\{1,-\tfrac{1}{3},-\tfrac{1}{3},-\tfrac{1}{3}\}$
(counted with algebraic multiplicity) and
$\mathrm{spec}(\mathcal{E}_2)=\{1,-1,0,0\}$.
Hence the peripheral eigenvalues are $\{1\}$ for $\mathcal{E}_1$ and $\{1,-1\}$ for $\mathcal{E}_2$,
i.\,e.\@, roots of unity as guaranteed by Lemma~\ref{lem:cyclic}.
It directly explains the eventual oscillation pattern contributed by $\mathcal{E}_2$. \qed
\end{example}

\section{Formal Logic Specification}\label{sect:logic}

Matrix Product State (MPS) is a popular formalism for representing quantum states in many-body systems.
Traditional temporal logics such as Linear Temporal Logic (LTL)
have been adapted to reason about properties of MPS
by interpreting the chain length as a temporal index.
However, this is a conceptual mismatch:
in MPS, we reason about spatial properties along a one-dimensional chain, not temporal evolution.
This motivates the introduction of \emph{Linear Chain Logic (LCL)}:
a logic designed specifically to express and verify properties of quantum states along spatial chains.

\subsection{Syntax and semantics of LCL}
Let $\text{val}_N$ denote a value (to be determined in detail below) extracted from
the MPS $\ket{\psi_N}$ at chain length $N$.
We start with basic properties of MPS that should be admitted by the logic:
\begin{itemize}
  \item $[\text{val}_N > \varepsilon]$ : the value at the current MPS exceeds a threshold,
  \item $[|\text{val}_N-\text{val}_1| < \delta]$ : correlation with the initial MPS is small,
  \item $[\frac{\text{val}_{N+1}}{\text{val}_{N}} < \gamma ]$ : exponential decay rate.
\end{itemize}
The first property usually needs to decide
whether the current energy $\|\psi_N\|^2 \linebreak[0] =\braket{\psi_N}{\psi_N}$ of MPS $\ket{\psi_N}$
meets an interval predicate $\in \mathcal{I}$;
if true, a corresponding label should be assigned to the current MPS $\ket{\psi_N}$.
For this end, we introduce the \emph{value expression} $\braket{\psi_N}{\psi_N}$
and the \emph{atomic formula} $\braket{\psi_N}{\psi_N} \in \mathcal{I}$
as the continuous-time linear logic (CLL)~\cite{guan2022probabilistic}.
The second property compares the current MPS $\ket{\psi_N}$ with the past one $\ket{\psi_1}$
in terms of energy.
So we need to pre-compute the constant value $\braket{\psi_1}{\psi_1}$ for the use afterwards.
The last property proposes the comparison
between an expression over the energies of the MPS and a threshold,
thus the value expression should be correspondingly generalised to admit:
i)~a linear expression and even a ratio of them;
ii)~the energies of the current MPS, the next ones,
and the initial ones involved as the variables in those expressions.
Unlike classic first-order logics where predicate and atomic formula are the same thing,
we design them in two layers ---
\emph{interval predicates} are the preparatory construct prior to \emph{atomic formulas} in LCL.
Such a syntactical separation allows us to specify the properties of MPS precisely.

\begin{example}\label{ex4}
    For the MPS $\ket{\psi_N}$ defined in Example~\ref{ex1},
    we consider the quantitative characterisation of its squared norm,
    namely $\Gamma_\textup{sqr}(N) = \braket{\psi_N}{\psi_N}$,
    which is a \emph{value expression} introduced here.
    Other value expressions like $\Gamma_\textup{dif}(N) = \braket{\psi_N}{\psi_N} - \braket{\psi_1}{\psi_1}$
    and $\Gamma_\textup{rat}(N) = \braket{\psi_{N+1}}{\psi_{N+1}}/\braket{\psi_N}{\psi_N}$
    are introduced similarly.
    Let $\mathcal{I}_\textup{pos}$ denote the positive interval $(0, \infty)$.
    We introduce the label $\ell_\textup{sqr,pos}$ to represent the \emph{atomic formula}
    that the current value expression falls into the positive interval,
    such that the label $\ell_\textup{sqr,pos}$ is assigned to the MPS $\ket{\psi_N}$
    if and only if $\Gamma_\textup{sqr}(N) \in \mathcal{I}_\textup{pos}$. \qed
\end{example}

To specify the essential properties mentioned above,
we propose the formal definition of LCL as follows.
\begin{definition}[Syntax]
An LCL formula $\phi$ is finitely generated by the syntax:
\begin{align*}
  \Phi \ \Coloneqq \ & \text{true} \mid \ell \mid \neg \Phi \mid \Phi_1 \wedge \Phi_2
  \mid \mathsf{X}\,\Phi \mid \mathsf{E}\,\Phi \mid \mathsf{G}\,\Phi
\end{align*}
where $\ell$ is a label representing some atomic formula over site-indexed observables or quantities,
and $\mathsf{X}$ (next), $\mathsf{E}$ (eventually), $\mathsf{G}$ (globally) are
the spatial operator to be interpreted along the chain.
\end{definition}
Here we do not include the until formula $\Phi_1\,\mathsf{U}\,\Phi_2$ as the standard LTL formulas,
because of simplification, no technical hardness.
In contrast to the until formula,
as the MPS are usually investigated in the limit behavior $N \to \infty$,
we should pay less attention to those non-ultimate states.
Hence the composite formula $\mathsf{E}\,\mathsf{G}\,\Phi$
would be of particular interest in verifying MPS.

As our intention that the LCL is devised for MPS,
we need a transformer from atomic formulas to labels,
which is constructed as follows.
For the current MPS $\ket{\psi_N}$, 
if a value expression $\Gamma_s$ on $\ket{\psi_N}$ falls into a range (interval) $\mathcal{I}_t$,
i.\,e.\@ $\Gamma_s(N) \in \mathcal{I}_t$,
a corresponding label $\ell_{s,t}$ would be assigned to this $\ket{\psi_N}$.
Namely, we interpret LCL on the chain model $\mathcal{C} = (S, \mathcal{V},L)$ of MPS, where:
\begin{itemize}
  \item $S = \setnocond{\ket{\psi_N}}$ is a (possibly infinite) sequence of chain lengths,
  \item $\mathcal{V}$ is the set of all labels $\ell$
  used to represent distinct value expressions falling into distinct intervals,
  \item $L: S \to \mathcal{V}$ is a labeling function that maps each chain length to labels.
\end{itemize}
In this model, $(\mathcal{C}, N)$ denotes the MPS $\ket{\psi_N}$ of length $N$ in $S$.
Then, we have:
\begin{definition}[Semantics]
    Under the chain model $\mathcal{C} = (S, \mathcal{V},L)$,
    the semantics of LCL is given by the satisfaction relation $(\mathcal{C}, N) \models \Phi$ as follows:
\begin{align*}
  (\mathcal{C}, N) &\models \text{true} && \text{always holds} \\
  (\mathcal{C}, N) &\models \ell && \text{if } \ell \in L(\ket{\psi_N}) \\
  (\mathcal{C}, N) &\models \neg \Phi && \text{if } (\mathcal{C}, N) \not\models \Phi \\
  (\mathcal{C}, N) &\models \Phi_1 \wedge \Phi_2 && \text{if }(\mathcal{C}, N) \models \Phi_1\text{ and }(\mathcal{C}, N) \models \Phi_2 \\
  (\mathcal{C}, N) &\models \mathsf{X}\,\Phi && \text{if } (\mathcal{C}, N+1) \models \Phi \\
  (\mathcal{C}, N) &\models \mathsf{E}\,\Phi && \text{if } \exists\, j \ge N, (\mathcal{C}, j) \models \Phi \\
  (\mathcal{C}, N) &\models \mathsf{G}\,\Phi && \text{if } \forall\, j \ge N, (\mathcal{C}, j) \models \Phi.
\end{align*}
\end{definition}

\begin{example}
Consider the sequence $S$ of MPS $\ket{N}$ in Example~\ref{ex1}
and the label set $\mathcal{V}=\setnocond{\ell_\textup{sqr,zero},\ell_\textup{sqr,pos}}$ in Example~\ref{ex4}.
They are used to construct the chain model $\mathcal{C} = (S, \mathcal{V},L)$
with the labeling function $L$ from $S$ to $\mathcal{V}$.
We have seen that $|\psi_{1}\rangle = 0$ and $|\psi_{2}\rangle \neq 0$.
Consequently, under the chain model $\mathcal{C}$,
we have $(\mathcal{C}, 1) \models \ell_\textup{sqr,zero}$
and $(\mathcal{C}, 2) \models \ell_\textup{sqr,pos}$,
as well as $(\mathcal{C}, 1) \models \mathsf{X}\,\ell_\textup{sqr,pos}$. \qed
\end{example}

Actually, we notice that
the truth of LCL formulas relies on the notion:
\begin{definition}[Evidence Set]
    Under a chain model $\mathcal{C}$,
    the evidence set of an LCL formula $\Phi$ is the set $\setcond{N}{(\mathcal{C}, N) \models \Phi}$.
\end{definition}
Once we obtain those evidence sets, deciding LCL formulas can be reduced to set operation,
which will be developed in the next section.

\subsection{Expressivity of LCL: more physically meaningful properties}\label{Subsec:properties}
Using LCL, we can express a rich class of physically meaningful properties as follows,
which will be composed to complicated ones for checking in Section~\ref{sect:exper}.

\begin{enumerate}
  \item \textbf{State validity:}
  $\mathsf{G}\,(\text{val}_N > 0)$
  with $\text{val}_N$ denoting the energy $\braket{\psi_N}{\psi_N}$ (and the same below).
  It exactly addresses the dichotomy problem (Problem~\ref{pro:many_body}).
  \item \textbf{Energy bounds:}
  $\mathsf{G}\,(\text{val}_N \in (1-\varepsilon,1+\varepsilon))$,
  where $\varepsilon$ is the perturbation tolerance.
  It concerns the stability of the CPTP map induced by some local MPS tensors.
  \item \textbf{Finite correlation size:}
  $\mathsf{E}\,\mathsf{G}\,(|\text{corr}_{1,N}| < \varepsilon)$
  with $\text{corr}_{1,N}$ denoting the difference $\text{val}_N-\text{val}_1$.
  It asks whether the energy of the MPS
  would have a difference less than $\varepsilon$ with the initial energy eventually always.
  \item \textbf{Ratio decay:}
  $\mathsf{G}\,(| \text{val}_{N+1}/\text{val}_N | < \gamma )$.
  It asks whether the energy of the MPS would always decay exponentially at a rate less than $\gamma$.
  \item \textbf{Oscillatory correlation:}
  $\mathsf{G}\,\big(\text{sign}(\text{corr}_{1,N}) \ne \text{sign}(\text{corr}_{1,N+1})\big)$.
  It asks if the energy of the MPS would always fluctuate around the initial energy.
  \item \textbf{Periodicity:}
  $\mathsf{G}\,(\text{val}_N = \text{val}_{N+k})$,
  where $k\in\mathbb N$ is a fixed period.
  It expects that the energy of the MPS would have a fixed period of $k$.
  \item \textbf{Phase transition indicator:}
  $\mathsf{E}\,\mathsf{G}\,(|\text{corr}_{i,j}| > \delta)$,
  where $\text{corr}_{i,j}$ is a selected two-size correlation function. 
  It expects that the energy difference of two MPS with size gap $|j-i|$ would exceed a threshold $\delta$ eventually always.
\end{enumerate}
The properties considered in Problem~\ref{pro:many_body} can be specified by the LCL formulas: 
\[
    \mathsf{X}^{J-1}\, \mathsf{G}\, \ell_\textup{sqr,pos}
    \quad \text{and} \quad
    \mathsf{X}^{J-1}\, \mathsf{E}\,\mathsf{G}\, \ell_\textup{sqr,pos}.
\]
LCL captures spatial reasoning along quantum chains,
aligning better with physical intuition and tensor network semantics than traditional temporal logics.
It provides a formal language to verify decay properties, symmetry, validity, and large-size order in MPS.

\section{Model Checking Algorithms}\label{sect:algo}
In this section we present the algorithm for checking LCL formulas for MPS.
It attempts to compute the evidence set of a given LCL formula,
following a topdown fashion,
which in the end is reduced to the ground of computing the evidence sets of atomic formulas.
Mathematically, it amounts to the sign determination of exponential polynomials,
closely similar to the Skolem problem~\cite{HHH+05} and its variants
--- the positivity and the ultimate positivity problems. 
However, due to the hardness of the Skolem problems,
we approximate the evidence set of an atomic formula from above and from below using semilinear sets.
It further yields an over-approximation and an under-approximation of the evidence set of the given LCL formula.

\subsection{Computing evidence sets in a topdown fashion}

\begin{algorithm}[htb]
	\caption{Computing Evidence Sets}\label{alg:check}
	\begin{algorithmic}[1]
		\item[] $$\Omega^\pm \Leftarrow \textsf{Check-All}(\mathcal{C}, \Phi)$$
	\Require $\mathcal{C}$ is a chain model of MPS $\setnocond{\ket{\psi_N}}$,
		and $\Phi$ is an LCL formula to be checked;
	\Ensure $\Omega^\pm$ is a superset/subset of
    all chain lengths $N$, satisfying $(\mathcal{C}, N) \models \Phi$.
	\If {$\Phi = \text{true}$}
		\State \Return $[0,\infty)$;
    \ElsIf {$\Phi = \ell$}
		\State\label{ln:ground} \Return $\textsf{Solve-Atomic}(\setnocond{\ket{\psi_N}}, \ell)$;
        \Comment{invoke Algorithm~\ref{alg:atomic} as a subroutine}
    \ElsIf {$\Phi = \neg \Phi_1$}
        \State compute $\Omega_1^\pm \Leftarrow \textsf{Check-All}(\mathcal{C}, \Phi_1)$ in a recursive fashion;
		\State \Return $[0,\infty) \setminus \Omega_1^\mp$;
        \Comment{switch the two approximations}
	\ElsIf {$\Phi = \Phi_1 \wedge \Phi_2$}
        \State compute $\Omega_1^\pm \Leftarrow \textsf{Check-All}(\mathcal{C}, \Phi_1)$
        and $\Omega_2^\pm \Leftarrow \textsf{Check-All}(\mathcal{C}, \Phi_2)$;
		\State \Return $\Omega_1^\pm \cap \Omega_2^\pm$;
    \ElsIf {$\Phi= \mathsf{X}\, \Phi_1$}
        \State compute $\Omega_1^\pm \Leftarrow \textsf{Check-All}(\mathcal{C}, \Phi_1)$;
        \State \Return $(\Omega_1^\pm \setminus \setnocond{0}) \ominus 1$;
        \Comment{$\ominus$ is element-wise minus, e.\,g.\@ $\setnocond{1,3}\ominus 1=\setnocond{0,2}$}
	\ElsIf {$\Phi= \mathsf{E}\,\Phi_1$}
        \State\label{ln:exist} compute $\Omega_1^\pm \Leftarrow \textsf{Check-All}(\mathcal{C}, \Phi_1)$;
        \If {$\Omega_1^\pm$ is not a finite set}\label{ln:exist1}
            \Return $[0,\infty)$;
        \ElsIf {$\Omega_1^\pm$ is not an empty set}\label{ln:exist2}
            \Return $[0,\sup \Omega_1^\pm]$;
        \Else\ \Return $\emptyset$;
        \EndIf
	\ElsIf {$\Phi= \mathsf{G}\,\Phi_1$}
        \State compute $\Omega_1^\pm \Leftarrow \textsf{Check-All}(\mathcal{C}, \mathsf{E}\,\neg\Phi_1)$;
        \State \Return $[0,\infty)\setminus \Omega_1^\mp$.
	\EndIf
\end{algorithmic}
\end{algorithm}

In Algorithm~\ref{alg:check},
if the ground of computing the evidence sets of an atomic formula (see Line~\ref{ln:ground}) gives an exact result,
the output of the whole algorithm would also be an exact result,
i.\,e.\@ $\Omega^+=\Omega^-$.
The correctness of Algorithm~\ref{alg:check} follows from the semantics of LCL,
but the computability, meaning that each instruction can be algorithmically realised,
relies on the observation (formulated below) that
the evidence set can be approximated by the structure of semilinear sets.
Recall that a \emph{semilinear} set $\Lambda$ is a finite set plus a finite union of arithmetic progressions.
Without loss of the generality, it is supposed to be put in the canonical form
$\Lambda = \Lambda_0 \uplus \biguplus_{\lambda \in \Lambda_1} (\lambda+\kappa\cdot\mathbb{N})$,
where $\Lambda_0,\Lambda_1$ are finite sets and $\kappa$ is the common difference,
satisfying $|\Lambda_1| \le \kappa$,
and $\mathbb{N}=\setnocond{0,1,2,\ldots}$.

\begin{lemma}[Approximate structure]\label{lem:struct}
The evidence set of an atomic formula in LCL can be over- and under-approximated by semilinear sets respectively.
\end{lemma}
From intuition, the proof idea can be explained as follows.
The value expression $\Gamma$ on MPS $\setnocond{\ket{\psi_N}}_{N \ge 1}$ defines a scalar sequence with
index $N$.
Whether its elements falling into a given interval $\mathcal{I}$ can be reduced to the sign determination problem.
Extending the sieve method,
we split the sequence into finitely many arithmetic progressions, expecting that:
i)~the sign of each arithmetic progression is exponentially stable in the \emph{ultimate};
ii)~there exists a \emph{computable} threshold,
after which the sign of the arithmetic progression is exponentially stable.
If so, the elements after that threshold are sign-invariant;
if they fulfill the sign condition,
all of them are contributed to some arithmetic progression $\lambda+\kappa\cdot\mathbb{N}$;
the elements before that threshold need to be checked one by one,
which are contributed to the finite set $\Lambda_0$.
Otherwise, the arithmetic progression has complicated sign condition,
which is \emph{indecisive}.
We include it by the over-approximation but exclude by the under-approximation.

We have seen that
the evidence set of an LCL formula can be approximated by the structure of semilinear sets,
since the family of semilinear sets is closed
under the set operations --- union, intersection and complement.
In Algorithm~\ref{alg:check}, manipulating semilinear sets are conducted on
finite sets $\Lambda_0,\Lambda_1$ and common difference $\kappa$,
which is standard in set theory.

\begin{example}\label{ex6}
Here we consider the LCL formula
$\Phi \coloneqq \mathsf{G}\,(\ell \rightarrow \mathsf{X}\,\ell)$,
in which the evidence set $\Omega(\ell)$ for $\ell$ is assumed to be
under- and over-approximated by $5+2\cdot\mathbb{N}$ and $1+2\cdot\mathbb{N}$ respectively.
(In fact, the two approximations will be obtained by the next example.)
Algorithm~1 evaluates $\Phi$ by recursively combining the semilinear
over- and under-approximations of $\Omega(\ell)$ through Boolean connective $\rightarrow$ (a syntactic sugar)
and spatial operators $\mathsf{X}$ and $\mathsf{G}$ appearing in $\Phi$ as follows:
\begin{enumerate}
    \item the evidence set of $\mathsf{X}\,\ell$ is approximated by $4+2\cdot\mathbb{N}$ and $2\cdot\mathbb{N}$,
    \item the evidence set of $\ell \rightarrow \mathsf{X}\,\ell$ is approximated by $2\cdot\mathbb{N}$
    and $\{1,3\}\cup 2\cdot\mathbb{N}$,
    \item the evidence set of $\Phi$ is approximated by $\Omega^+(\Phi)=2\cdot\mathbb{N}$
    and $\Omega^-(\Phi)=2\cdot\mathbb{N}$,
\end{enumerate}
yielding the corresponding evidence set of $\Phi$ is $\Omega(\Phi)=2\cdot\mathbb{N}$. \qed
\end{example}

\subsection{Approximately computing evidences of an atomic formula}

\begin{algorithm}[htb]
	\caption{Approximately Computing Evidences of an Atomic Formula}\label{alg:atomic}
	\begin{algorithmic}[1]
		\item[] $$\Omega^\pm \Leftarrow \textsf{Solve-Atomic}(\setnocond{\ket{\psi_N}}, \ell)$$
	\Require $\setnocond{\ket{\psi_N}}$ is the MPS
        defined by a family of $D$-by-$D$ matrices $\setnocond{A_k}_{k=1}^d$,
		and $\ell$ is a label corresponding the atomic formula $\Gamma(N) \in \mathcal{I}$;
	\Ensure $\Omega^\pm$ is a superset/subset of all chain lengths $N$,
        satisfying $\ket{\psi_N} \models \ell$.
    \State let $\e=\setnocond{A_k}_{k=1}^d$ be a super-operator, and $\Omega^\pm$ be initialised as $\emptyset$;
    \State compute an irreducible form $(\h_m,\e_m)$, from $\e$, with spectral radius $r_m$;
    \State let $p_m$ be the period of the peripheral eigenvalues of $\e_m$,
    and $\kappa=\mathrm{lcm}(\setnocond{p_m})$;
    \State let $\Gamma|_r$ denote the value expression $\Gamma$ projected with the peripheral eigenvalues
    from irreducible forms $(\h_m,\e_m)$ whose spectral radius $r_m=r$;
    \ForAll {$1 \le i \le \kappa$}\label{ln:enumerate} \Comment{check each residue class}
        \If{there is a spectral radius $r$ such that $\Gamma|_r(i) \ne 0$}
            \State let $r$ be the maximum of such a radius;
            \If{$\Gamma|_r(N)$ is an interior point of $\mathcal{I}$
            for any large $N \equiv i$ ($\mathrm{mod}\,\kappa$)}\label{ln:cond}
                \If{$r$ is greater than the 2nd spectral radius of $\e_m$ for all $r_m>r$}\label{ln:cond1}
                    \State compute a threshold $N_i \equiv i$
                    such that $\Gamma(j) \in \mathcal{I}$ ($j\in N_i+\kappa\cdot \mathbb{N}$);
                    \State find all small integers $j \equiv i$, counted in $\Lambda_i$,
                    for which $\ket{\psi_j} \models \ell$ holds;
                    \State update $\Omega^- \gets \Omega^- \uplus \Lambda_i \uplus (N_i+\kappa\cdot \mathbb{N})$;
                \EndIf
                \State update $\Omega^+ \gets \Omega^+ \uplus (i+\kappa\cdot \mathbb{N})$;
            \EndIf
        \EndIf
    \EndFor
    \State \Return $\Omega^\pm$.
\end{algorithmic}
\end{algorithm}

In Algorithm~\ref{alg:atomic},
the spectral radius $r_m$ of $\e_m$ can be computed as that of the matrix representation $M_{\e_m}$ of $\e_m$.
We numerically compute $\tr(M_{\e_m}^N)$ for a few successive large values of $N$,
which turns out to be periodic nearly with the pattern
$\underbrace{0,\ldots,0}_{p_m-1\text{ copies}},p_m r_m^N$,
since i)~peripheral eigenvalues are dominant,
and ii)~they are the spectral radius multiplied with roots of unity by Lemma~\ref{lem:cyclic}.
Thus the period $p_m$ is acquired as well as the least common period $\kappa$.
The peripheral eigenvalues $\lambda_{m,1},\ldots,\lambda_{m,p_m}$ can further be determined
as $r_m$ multiplied with $p_m$th roots of unity,
as well as the corresponding eigenvectors  $\ket{\lambda_{m,1}},\ldots,\ket{\lambda_{m,p_m}}$.
We claim that the second spectral radius of $\e_m$ is the spectral radius of the matrix
\begin{equation}\label{eq:difference}
    M_{\e_m} - \sum_{l=1}^{p_m} \lambda_{m,l}\ketbra{\lambda_{m,l}}{\lambda_{m,l}}.
\end{equation}
This is because: i)~peripheral eigenvalues have algebraic multiplicity 1
and their eigenvectors are orthogonal to the eigen-subspaces of other eigenvalues;
ii)~the blocks associated with peripheral eigenvalues in the Jordan decomposition of $M_{\e_m}$
are removed exactly by deleting $\sum_{l=1}^{p_m} \lambda_{m,l}\ketbra{\lambda_{m,l}}{\lambda_{m,l}}$.

Based on the preparation, we explain the rationale of Algorithm~\ref{alg:atomic} as follows.
The value expression $\Gamma(N)$ is eventually $\kappa$-periodic.
In Line~\ref{ln:enumerate},
we step into each residue class ---
the arithmetic progression $\setnocond{\Gamma(N)}_{N \equiv i \ (\mathrm{mod}\,\kappa)}$.
If the condition in Line~\ref{ln:cond} is satisfied,
the current arithmetic progression is a candidate to fulfill the sign condition
extracted from $\Gamma(N) \in \mathcal{I}$.
Further, if the condition in Lines~\ref{ln:cond1} is satisfied,
there is the suffix of the current arithmetic progression fulfills that sign condition;
in this case the increment in $\Omega^+$ and $\Omega^-$ can differ only on a finite prefix of that progression,
which ensures a small gap between $\Omega^+$ and $\Omega^-$.

\begin{example}\label{ex7}
We resume Example~\ref{ex3} with the irreducible form $(\h_m,\e_m)$ ($m=1,2$),
and consider the atomic formula
$\ell \equiv [\Gamma(N)\in \mathcal{I}]$,
where $\Gamma(N)=\braket{\psi_N}{\psi_N}$ and $\mathcal{I}=(0.95,1.05)$.
For $\mathcal{E}_1$ and $\mathcal{E}_2$,
their spectral radii are both $\rho=1$ with periods $p_1=1$ and $p_2=2$ respectively.
Accordingly, we split the sequence $\setnocond{\Gamma(N)}_{N \in \mathbb{N}}$
into the odd subsequence $\setnocond{\Gamma(2N-1)}$ and the even one $\setnocond{\Gamma(2N)}$,
as $\kappa=\mathrm{lcm}(p_1,p_2)=2$.

For details, we can deliver the peripheral eigenspaces explicitly.
The channel $\e_1$ has the spectral radius $1$
with the normalised eigenvector $\gamma_1=\tfrac{1}{\sqrt{2}}I$,
i.\,e.\@ $\e_1(\gamma_1)=\gamma_1$.
After vectorising $\gamma_1$ as a column vector $\ket{\lambda_1} \in \mathbb{C}^4$,
the second spectral radius of $\e_1$ is computed as
the spectral radius $\tfrac{1}{3}$ of $M_{\e_1}-\ketbra{\lambda_1}{\lambda_1}$.
Similarly, $\e_2$ has the second spectral radius 0.

Next, according to~\eqref{eq:trace-spectral-sum}, we have the interval estimate:
\[
  \Gamma(N)\in \Big[\underbrace{1^N+1^N+(-1)^N}_{\text{peripheral eigenvalues}} - \frac{\color{red}3}{3^N},\
  \underbrace{1^N+1^N+(-1)^N}_{\text{peripheral eigenvalues}} + \frac{\color{red}3}{3^N} \Big],
\]
where the numerator $\color{red}3$ comes from the number of nonperipheral eigenvalues of $\e_1$.
Restricting to the even subsequence gives $\Gamma(2N) \ge \tfrac{8}{3}$ for all $N$,
hence $\Gamma(2N)\notin \mathcal{I}$;
$\Gamma(2N-1)$ tends to 1,
so only the odd subsequence can eventually satisfy $\ell$.
Algorithm~\ref{alg:atomic} returns the over-approximation $\Omega^+(\ell)=1+2\cdot\mathbb{N}$,
and it computes a (tight) threshold $N_1=5$ such that $\Gamma(2N+1) \in \mathcal{I}$ for all $2N+1\ge N_1$.
Similarly, we get $\Omega^-(\ell)=5+2\cdot\mathbb{N}$,
which has already been used in Example~\ref{ex6}.
It has been validated that the initial values are $\Gamma(1)=0$,
$\Gamma(3)=8/9$, $\Gamma(5)=80/81$, $\Gamma(7)=728/729$ and thus always fall into the interval $\mathcal{I}$;
while even $N$ yields $\Gamma(2)=10/3$, $\Gamma(4)=82/27$, $\Gamma(6)=730/243$,
thus never falling into $\mathcal{I}$. \qed
\end{example}

We turn to the complexity issue.
Algorithm~\ref{alg:check} plays the role of a high-level traverse for the syntactic tree of $\Phi$;
Algorithm~\ref{alg:atomic} is the low-level treatment for each leaf in that tree.
The irreducible forms in Line~2 of Algorithm~\ref{alg:atomic} can be obtained in time $\mathcal{O}(D^8)$
as computing bottom strongly connected components in a QMC~\cite[Theorem~17]{feng2017model}.
The second spectral radius in Line~9 can also be computed in time $\mathcal{O}(D^8)$ by
obtaining all eigenvectors associated with peripheral eigenvalues in Eq.~\eqref{eq:difference}.
All other operations are not expensive.
Hence, we have:
\begin{theorem}
    Algorithm~\ref{alg:check} incorporated with Algorithm~\ref{alg:atomic} is in time $\mathcal{O}(|\Phi| \cdot D^8)$.
\end{theorem}

\section{Experimentation}\label{sect:exper}
This section delivers experiments showing the scalability of the proposed method,
and the diversity when meeting different models and complicated properties.
Focusing on computational performance,
we check and report all pairs of models and properties.
Physically meaningful discussions bridging the models with their condensed-matter physics guarantees are provided in Appendix~\ref{sect:pairs}.

\subsection{Benchmarks and model preparation}
We use two benchmark suites:
\emph{(A) synthetic scalable channel families} to stress-test scalability
and \emph{(B) physical spin-chains} to show relevance to standard many-body models.
Synthetic families scale up to bond dimension $D=128$,
while physical models are tested at $D\le 32$.
Please find the details in Appendix~\ref{sec:exp-models}.

\paragraph{(A) Synthetic scalable channels (stress tests).}
Each family starts from a small $2\times2$ CPTP map given by Kraus operators $\{K_i\}$ and is lifted to
bond dimension $D=2^t$ by tensor power
$K_{(i_1,\ldots,i_t)} \coloneqq K_{i_1}\otimes\cdots\otimes K_{i_t}$.
We include five representative behaviors:
\textit{AKLT-like} (fast-mixing)~\cite{Affleck1987},
\textit{cluster-like} (highly structured mixing)~\cite{Raussendorf2001},
\textit{random-gapped} (controlled Pauli-noise gap)~\cite{GonzalezGuillen2018},
\textit{near-critical} (slow-mixing hard instances)~\cite{HastingsKoma2006},
and \textit{periodic} (non-primitive transfer inducing oscillations)~\cite{Perez-Garcia2007}.

\paragraph{(B) Physical spin-chains (realistic cases).}
We test three standard 1D models and use their translation-invariant ground states in MPS form:
\textit{TFIM}~\cite{Pfeuty1970},
\textit{XXZ}~\cite{Giamarchi2003}, and \textit{Kitaev chain}~\cite{Kitaev2001}.
For each, we obtain ground-state tensors $\{A_1,A_2\}\subset\mathbb{C}^{D\times D}$
with a standard infinite-system TN routine (e.\,g.\@, iDMRG/VUMPS) in representative near-critical regimes,
and gauge-fix to a stable canonical form.

\subsection{LCL specifications}\label{subs:spec}
By composing the seven basic LCL formulas listed in Subsection~\ref{Subsec:properties},
we evaluate the following complicated properties for demonstrating efficiency:
\begin{align*}
  \Phi_1 &\coloneqq \mathsf{X}^{J-1}\,\mathsf{G}\,\ell_{\mathrm{nz}}
    &&\text{ with }  \ell_{\mathrm{nz}} \equiv \text{val}_N > 0, \\
  \Phi_2 &\coloneqq \mathsf{X}^{J-1}\,\mathsf{G}\,\ell_{\mathrm{bd}}
    &&\text{ with } \ell_{\mathrm{bd}} \equiv \text{val}_N \in (1-\varepsilon,\,1+\varepsilon), \\
  \Phi_2' &\coloneqq \mathsf{X}^{J-1}\,\mathsf{G}\,(\ell_{\mathrm{bd}} \wedge \ell_{\mathrm{clust}})
    &&\text{ with } \ell_{\mathrm{clust}} \equiv |\text{corr}_{1,N}| < \varepsilon, \\
  \Phi_3 &\coloneqq \mathsf{E}\,\mathsf{G}\,(\ell_{\mathrm{rat}} \wedge \mathsf{X}\,\ell_{\mathrm{rat}})
    &&\text{ with } \ell_{\mathrm{rat}} \equiv \left|\text{val}_{N+1}/\text{val}_N\right| < \gamma, \\
  \Phi_4 &\coloneqq \mathsf{X}^{J-1}\,\mathsf{G}\,(\ell_{\mathrm{clust}} \wedge \neg \ell_{\mathrm{osc}})
    &&\text{ with } \ell_{\mathrm{osc}} \equiv \mathrm{sign}(\text{corr}_{1,N}) \neq \mathrm{sign}(\text{corr}_{1,N+1}), \\
  \Phi_5 &\coloneqq \mathsf{X}^{J-1}\,\mathsf{G}\,(\neg \ell_{\mathrm{per},k})
    &&\text{ with } \ell_{\mathrm{per},k} \equiv \text{val}_N = \text{val}_{N+k}, \\
  \Phi_6 &\coloneqq \mathsf{E}\,\mathsf{G}\,(\ell_{\mathrm{lro}} \wedge \neg \ell_{\mathrm{clust}})
    &&\text{ with } \ell_{\mathrm{lro}} \equiv |\text{corr}_{i,j}| > \delta .
\end{align*}
Here, $\Phi_1$ is specified for the property of stable state validity,
$\Phi_2$ is for preservation of normalisation,
$\Phi_2'$ is for joint normalization and clustering,
$\Phi_3$ is for stable ratio decay,
$\Phi_4$ is for pure clustering without oscillations,
$\Phi_5$ is for asymptotic aperiodicity (not $k$-periodic),
and $\Phi_6$ is for large-size order without clustering.
The parameters are chosen as $\varepsilon=0.01$, $\gamma=1.0$, $\delta=0.01$, $k=2$ and $J=1$ in our experiments.

\subsection{Experimental results}
We experiment on an Apple Silicon M4 Mac running macOS 26.
The machine is equipped with a 10-core CPU and 10-core GPU sharing 24~GiB of unified memory.
We use the MLX and NumPy Python packages to compute the required numerical quantities;
matrix operations are dispatched to the Apple Metal GPU backend via MLX.

Tables~\ref{tab:synthetic} and~\ref{tab:physical} show the results on
(A)~synthetic scalable channel families and (B)~physical spin-chain families, respectively.
In the tables, each evaluated $(\text{model},\Phi,D)$ instance is reported in the form
\[
    \texttt{Verdict / Runtime / PeakMemory}.
\]
The symbols are explained as follows.
\begin{itemize}
  \item \textbf{T (True):} the formula $\Phi$ is verified to be true.
  \item \textbf{F (False):} the formula $\Phi$ is verified to be false.
  \item \textbf{U (Unknown):} the algorithm terminates,
  but the truth of $\Phi$ cannot be determined due to the approximation.
  \item \textbf{TO (Timeout):} the computation exceeds the time limit of 3600 seconds.
\end{itemize}
For entries reported as \texttt{U / time / memory}, the algorithm terminates but yields an unknown verdict.
For entries reported as \texttt{-- / TO / --},
the algorithm does not complete and no semantic conclusion can be drawn.
In particular, all 42 instances in Table~\ref{tab:physical} are completed within the 3600-second timeout.

\begin{sidewaystable*}[htbp]
\centering
\caption{Synthetic scalable channel families: decision (\textbf{T}/\textbf{F}/\textbf{U}/\textbf{TO}),
total runtime (seconds), and peak memory (MB) for each $(\text{model},\,\Phi,\,D)$ instance.}
\label{tab:synthetic}
\scriptsize
\setlength{\tabcolsep}{3.5pt}
\renewcommand{\arraystretch}{1.1}
\begin{tabular}{llcccc}
\toprule
\multicolumn{2}{c}{} & \multicolumn{4}{c}{Bond dimension $D$} \\
\cmidrule(lr){3-6}
Model & Formula & 16 & 32 & 64 & 128 \\
\midrule
\multirow{7}{*}{AKLT-like}
  & $\Phi_1$ & T / 0.0655 / 1.955  & T / 0.9952 / 25.73  & T / 1.2700 / 241.7  & T / 7.1690 / $2.917{\times}10^{3}$ \\
  & $\Phi_2$ & T / 0.0495 / 1.953  & T / 0.3983 / 25.73  & T / 0.9443 / 241.7  & T / 11.510 / $2.917{\times}10^{3}$ \\
  & $\Phi_2'$ & F / 0.0465 / 1.952  & F / 0.2558 / 25.73  & F / 1.5360 / 241.7  & F / 8.8220 / $2.917{\times}10^{3}$ \\
  & $\Phi_3$ & F / 0.0830 / 1.953  & F / 0.4872 / 25.73  & F / 1.9350 / 241.7  & U / 9.0460 / $2.917{\times}10^{3}$ \\
  & $\Phi_4$ & F / 0.0568 / 1.952  & F / 0.3210 / 25.73  & F / 1.0400 / 241.7  & F / 8.4480 / $2.917{\times}10^{3}$ \\
  & $\Phi_5$ & F / 0.0350 / 1.952  & F / 0.6056 / 25.73  & F / 1.1260 / 241.7  & F / 8.5780 / $2.917{\times}10^{3}$ \\
  & $\Phi_6$ & T / 0.0382 / 1.952  & T / 0.2495 / 25.73  & T / 1.2900 / 241.7  & T / 8.7270 / $2.917{\times}10^{3}$ \\
\midrule
\multirow{7}{*}{Cluster-like}
  & $\Phi_1$ & T / 0.0179 / 0.166  & T / 0.0090 / 1.140   & T / 0.1053 / 8.525  & T / 0.1938 / 65.92 \\
  & $\Phi_2$ & T / 0.0149 / 0.166  & T / 0.0078 / 1.140   & T / 0.1104 / 8.525  & T / 0.1697 / 65.92 \\
  & $\Phi_2'$ & T / 0.0118 / 0.166  & T / 0.0064 / 1.139   & T / 0.1531 / 8.525  & T / 0.2755 / 65.93 \\
  & $\Phi_3$ & F / 0.0219 / 0.166  & F / 0.0083 / 1.140   & F / 0.1208 / 8.525  & T / 0.1716 / 65.93 \\
  & $\Phi_4$ & T / 0.0285 / 0.166  & T / 0.0097 / 1.140   & T / -0.0003 / 8.525 & T / 0.3378 / 65.92 \\
  & $\Phi_5$ & F / 0.0239 / 0.167  & F / 0.0099 / 1.140   & F / 0.1425 / 8.525  & F / 0.2175 / 65.93 \\
  & $\Phi_6$ & F / 0.0193 / 0.166  & F / 0.0082 / 1.139   & F / 0.1315 / 8.525  & F / 0.2213 / 65.92 \\
\midrule
\multirow{7}{*}{Near-critical}
  & $\Phi_1$ & T / 0.9421 / 2.997  & T / 0.6857 / 3.741  & U / 0.7700 / 24.73  & --- / TO / --- \\
  & $\Phi_2$ & F / 0.7970 / 0.167  & F / 0.6423 / 3.741  & F / 0.7102 / 24.73  & --- / TO / --- \\
  & $\Phi_2'$ & F / 0.5053 / 1.664  & F / 0.4307 / 3.741  & F / 0.4234 / 24.73  & --- / TO / --- \\
  & $\Phi_3$ & U / 1.3480 / 4.191  & U / 1.3030 / 4.643  & U / 1.0990 / 24.73  & U / 0.9794 / 163.6 \\
  & $\Phi_4$ & F / 1.3430 / 1.726  & F / 1.0610 / 3.741  & F / 1.1580 / 24.73  & F / 1.1410 / 163.6 \\
  & $\Phi_5$ & T / 1.0240 / 1.498  & T / 0.9007 / 3.742  & T / 1.0020 / 24.73  & T / 1.0410 / 163.6 \\
  & $\Phi_6$ & T / 0.9709 / 6.046  & T / 0.8496 / 3.742  & T / 0.7658 / 24.73  & --- / TO / --- \\
\midrule
\multirow{7}{*}{Periodic}
  & $\Phi_1$ & T / 0.0982 / 2.553  & T / 0.1753 / 30.00  & T / 0.7437 / 361.3  & T / 10.990 / $4.353{\times}10^{3}$ \\
  & $\Phi_2$ & T / 0.0986 / 2.553  & T / 0.1126 / 30.00  & T / 0.7650 / 361.3  & T / 10.900 / $4.353{\times}10^{3}$ \\
  & $\Phi_2'$ & F / 0.0949 / 2.553  & F / 0.1489 / 30.00  & F / 0.7042 / 361.3  & F / 11.000 / $4.353{\times}10^{3}$ \\
  & $\Phi_3$ & F / 0.1675 / 2.553  & F / 0.2147 / 30.00  & F / 0.8179 / 361.3  & U / 10.660 / $4.353{\times}10^{3}$ \\
  & $\Phi_4$ & F / 0.1608 / 2.552  & F / 0.2263 / 30.00  & F / 0.8019 / 361.3  & F / 10.480 / $4.353{\times}10^{3}$ \\
  & $\Phi_5$ & F / 0.1699 / 2.553  & F / 0.1954 / 30.00  & F / 0.7686 / 361.3  & F / 10.600 / $4.353{\times}10^{3}$ \\
  & $\Phi_6$ & T / 0.1457 / 2.553  & T / 0.1529 / 30.00  & T / 0.7424 / 361.3  & T / 10.890 / $4.353{\times}10^{3}$ \\
\midrule
\multirow{7}{*}{Random-gapped}
  & $\Phi_1$ & T / 0.0964 / 2.552  & T / 0.2067 / 30.00  & T / 1.1540 / 361.3  & T / 22.780 / $4.353{\times}10^{3}$ \\
  & $\Phi_2$ & T / 0.0844 / 2.552  & T / 0.3022 / 30.00  & T / 0.8070 / 361.3  & T / 33.740 / $4.353{\times}10^{3}$ \\
  & $\Phi_2'$ & F / 0.0730 / 2.552  & F / 0.2113 / 30.00  & F / 0.8240 / 361.3  & F / 15.940 / $4.353{\times}10^{3}$ \\
  & $\Phi_3$ & T / 0.1145 / 2.552  & T / 0.1927 / 30.00  & T / 1.2200 / 361.3  & T / 29.240 / $4.353{\times}10^{3}$ \\
  & $\Phi_4$ & F / 0.1212 / 2.552  & F / 0.2040 / 30.00  & F / 1.5570 / 361.3  & F / 21.590 / $4.353{\times}10^{3}$ \\
  & $\Phi_5$ & F / 0.1086 / 2.553  & F / 0.2254 / 30.00  & F / 1.1000 / 361.3  & F / 17.610 / $4.353{\times}10^{3}$ \\
  & $\Phi_6$ & T / 0.1007 / 2.552  & T / 0.1923 / 30.00  & T / 0.9489 / 361.3  & T / 16.350 / $4.353{\times}10^{3}$ \\
\bottomrule
\end{tabular}
\end{sidewaystable*}

\begin{table}[htb!]
\centering
\caption{Physical spin-chain families: decision (\textbf{T}/\textbf{F}/\textbf{U}/\textbf{TO}),
  total runtime (seconds), and peak memory (MB) for each $(\text{model},\,\Phi,\,D)$ instance.}
\label{tab:physical}
\scriptsize
\setlength{\tabcolsep}{5pt}
\renewcommand{\arraystretch}{1.1}
\begin{tabular}{llcc}
\toprule
\multicolumn{2}{c}{} & \multicolumn{2}{c}{Bond dimension $D$} \\
\cmidrule(lr){3-4}
Model & Formula & 16 & 32 \\
\midrule
\multirow{7}{*}{TFIM}
  & $\Phi_1$ & T / 0.4560 / 2.949  & T / 0.4899 / 2.953 \\
  & $\Phi_2$ & F / 0.4938 / 0.167  & T / 0.4922 / 4.123 \\
  & $\Phi_2'$ & F / 0.2982 / 0.167  & F / 0.3094 / 0.212 \\
  & $\Phi_3$ & T / 0.7437 / 2.949  & T / 0.7943 / 2.953 \\
  & $\Phi_4$ & F / 0.8185 / 0.167  & F / 0.8615 / 0.212 \\
  & $\Phi_5$ & F / 0.7181 / 1.333  & F / 0.7091 / 1.242 \\
  & $\Phi_6$ & T / 0.6153 / 2.948  & T / 0.6334 / 2.954 \\
\midrule
\multirow{7}{*}{XXZ}
  & $\Phi_1$ & T / 0.0023 / 0.043  & T / 0.0023 / 0.043 \\
  & $\Phi_2$ & T / 0.0023 / 0.065  & T / 0.0023 / 0.065 \\
  & $\Phi_2'$ & T / 0.0019 / 0.022  & T / 0.0018 / 0.022 \\
  & $\Phi_3$ & F / 0.0037 / 0.007  & F / 0.0036 / 0.007 \\
  & $\Phi_4$ & T / 0.0049 / 0.043  & T / 0.0050 / 0.043 \\
  & $\Phi_5$ & F / 0.0067 / 0.022  & F / 0.0052 / 0.022 \\
  & $\Phi_6$ & F / 0.0038 / 0.023  & F / 0.0040 / 0.023 \\
\midrule
\multirow{7}{*}{Kitaev chain}
  & $\Phi_1$ & T / 0.0023 / 0.036  & T / 0.0026 / 0.036 \\
  & $\Phi_2$ & T / 0.0023 / 0.036  & T / 0.0025 / 0.036 \\
  & $\Phi_2'$ & F / 0.0023 / 0.036  & F / 0.0023 / 0.036 \\
  & $\Phi_3$ & T / 0.0024 / 0.036  & T / 0.0027 / 0.036 \\
  & $\Phi_4$ & F / 0.0028 / 0.036  & F / 0.0031 / 0.036 \\
  & $\Phi_5$ & F / 0.0027 / 0.036  & F / 0.0028 / 0.037 \\
  & $\Phi_6$ & T / 0.0025 / 0.036  & T / 0.0028 / 0.036 \\
\bottomrule
\end{tabular}
\end{table}

\section{Conclusion}
We broadened quantum model checking from programs and protocols to 1D many-body physics,
where properties were naturally indexed by the chain length.
Our key bridge was the transfer-operator view: a periodic MPS induced a CP map,
turning MPS contractions into traces of matrix powers and reducing nonzeroness to a tractable scalar sequence.
On top of this, we introduced LCL for specifying size-indexed spatial properties
of MPS families and gave approximate model-checking algorithms with sound over/under-approximations.
Our experiments covered two sets of models:
i)~scalable synthetic channel families to stress-test large bond dimensions,
and ii)~ground-state MPS from standard spin-chains to reflect realistic physical instances.
Experimental results showed that the checker could efficiently certify many properties in gapped and mixing regimes,
and suggested that logic-based verification would be a practical complement
to traditional tensor-network analysis for many-body physics, like PEPS and MERA~\cite{cirac2021matrix}.

\newpage
\bibliographystyle{unsrt}
\bibliography{main-final}

\newpage
\appendix

\section{Supplementary Experimental Materials}
\subsection{Details on model preparation}\label{sec:exp-models}

We consider two families of models:
(A)~\emph{synthetic-but-structured} scalable channel families, and
(B)~\emph{physical} spin-chain families represented by their translation-invariant ground-state MPS tensors.
The two families serve distinct roles in our evaluation:
synthetic families are used to stress-test scalability up to large bond dimensions ($D=128$),
while physical spin-chain families are evaluated at moderate bond dimensions ($D\le 32$),
reflecting realistic limits of infinite-system tensor-network solvers near criticality.

\paragraph{(A) Synthetic scalable channel families.}
For each model family, we begin with a fixed set of $2\times2$ Kraus operators
$\{K_i\}_{i=1}^{r}$ defining a CPTP map.

\begin{itemize}
  \item \textbf{AKLT-like}
  is a gapped, fast-mixing proxy inspired by valence-bond constructions (finite correlation size).
  We use three Kraus operators
  \[
    K_{+}=\sqrt{\tfrac{2}{3}}\sigma^+,\qquad
    K_{0}=-\tfrac{1}{\sqrt3}Z,\qquad
    K_{-}=-\sqrt{\tfrac{2}{3}}\sigma^-,
  \]
  where $\sigma^\pm=\tfrac{1}{2}(X\pm i Y)$.

  \item \textbf{Cluster-like}
  is a stabilizer/Clifford-like proxy exhibiting structured mixing
  (useful for bounded observables and oscillation tests):
  \[
    K_1=\tfrac{1}{\sqrt2}\mathbf{H},\qquad
    K_2=\tfrac{1}{\sqrt2}\mathbf{H}Z,
  \]
  where $\mathbf{H}$ is the Hadamard matrix $(X+Z)/{\sqrt2}$.
  \item \textbf{Random-gapped}
  is a controllable gapped family defined by Pauli noise:
  \[
    K_0=\sqrt{1-2\epsilon_5}\,I,\qquad
    K_1=\sqrt \epsilon_5\,X,\qquad
    K_2=\sqrt \epsilon_5\,Z,
  \]
  with a fixed $\epsilon_5=0.05$.

  \item \textbf{Near-critical}
  is a slow-mixing proxy with a small effective gap (hard instances; large unknown zones):
  \[
    K_1=\sqrt{1-\epsilon_1}\,\mathrm{diag}(1,e^{i\theta}),\qquad
    K_2=\sqrt{\epsilon_1}\,Z,
  \]
  with $\theta=0.98\pi$ and $\epsilon_1=0.01$.

  \item \textbf{Periodic}
  is a non-primitive periodic transfer structure inducing oscillatory correlations:
  \[
    U=\mathrm{diag}(1,i),\quad
    K_1=\sqrt{1-\epsilon_2}\,U,\quad
    K_2=\sqrt{\tfrac{\epsilon_2}{2}}\,X,\quad
    K_3=\sqrt{\tfrac{\epsilon_2}{2}}\,Y,
  \]
  with $\epsilon_2=0.02$.
\end{itemize}
To scale from the base $2\times2$ channel to bond dimension $D=2^t$,
we take the self-tensor power of the Kraus operators:
\[
  K_{(i_1,\dots,i_t)} \coloneqq K_{i_1}\otimes\cdots\otimes K_{i_t}.
\]  

\paragraph{(B) Physical spin-chain families.}
We additionally include three standard many-body models
and use their translation-invariant ground states represented as MPS tensors.
For each physical model,
we choose bond dimensions $D\in\{16,32\}$,
which are representative of physically meaningful
and computationally feasible truncation levels for infinite-system tensor-network solvers.
The MPS tensors $\{A_1,A_2\}\subset\mathbb{C}^{D\times D}$
are obtained via a standard infinite-system TN routine (e.\,g.\@, iDMRG/VUMPS)
at representative parameter regimes (gapped and near-critical where applicable),
and then they are gauge-fixed to a stable canonical form.

\begin{itemize}
  \item \textbf{TFIM}
  is a transverse-field Ising model $H=-J\sum_i Z_i Z_{i+1}-h\sum_i X_i$,
  exhibiting gapped and critical phases depending on $h/J$,
  where the subscript $i$ labels the lattice site for which the Pauli operators $X, Z$ act on.

  \item \textbf{XXZ}
  is a spin-chain
  $H=\sum_i (S_i^x S_{i+1}^x + S_i^y S_{i+1}^y + \Delta S_i^z S_{i+1}^z)$,
  with both gapped and critical regimes,
  where $S_i^{x,y,z}$ denote the spin-$\tfrac12$ operators acting on site $i$.

  \item \textbf{Kitaev chain}
  is a 1D topological superconducting chain with gapped and near-critical parameter regimes,
  in which oscillatory correlations may arise.
\end{itemize}

\subsection{Physical meaning of LCL Properties}\label{sect:pairs}
This subsection provides the rigorous physical justification for the LCL properties raised in Subsections~\ref{Subsec:properties} and~\ref{subs:spec},
as well as the rationale behind their targeted pairing with the physical spin-chain families.
The expressivity of LCL allows for the direct translation of canonical phenomena in 1D quantum many-body physics
into formal, verifiable specifications.
\begin{itemize}
    \item \textbf{State validity and energy bounds ($\Phi_1$, $\Phi_2$):}
    The validity condition ($\text{val}_N > 0$) and normalization bounds
    reflect the foundational theory of finitely correlated states~\cite{fannes1992finitely}.
    The stability of these properties over the chain is governed by
    the spectral radius of the CPTP map induced by the local MPS tensors~\cite{Perez-Garcia2007,wolf2012quantum}.
    LCL formalises the verification of this transfer-operator stability.
    
    \item \textbf{Finite correlation size and ratio decay ($\Phi_2'$, $\Phi_3$, $\Phi_4$):}
    LCL properties evaluating exponential clustering (e.\,g.\@, $|\text{corr}_{1,N}| < \varepsilon$)
    directly formalise the rigorously proven theorem that
    a spectral gap implies exponential decay of correlations~\cite{HastingsKoma2006}.
    In the transfer-matrix formalism,
    the ratio decay evaluated by $\Phi_3$ is dictated by the physical correlation length $\xi = -1/\ln|\lambda_2/\lambda_1|$,
    where $\lambda_i$ are the leading eigenvalues~\cite{schollwock2011dmrg}.
    Verifying strict decay ensures the modeled system properly adheres to the 1D area law of entanglement~\cite{hastings2007area}.
    
    \item \textbf{Oscillatory correlation and periodicity ($\Phi_5$, $\neg\ell_{\mathrm{osc}}$):}
    Formulas probing asymptotic aperiodicity and oscillatory correlation detect non-primitive transfer matrices and complex peripheral eigenvalues. Physically, these conditions identify incommensurate phases, spiral correlations, or discrete translational symmetry breaking,
    such as dimerization~\cite{zauner2015transfer,haegeman2017diagonalizing}.
    
    \item \textbf{Phase transition indicators ($\Phi_6$):}
    By checking for large-size order without strict clustering, $\Phi_6$ serves as an indicator of phase transitions.
    Because continuous spontaneous symmetry breaking (SSB) is forbidden in 1D quantum systems at finite temperature \cite{mermin1966absence},
    this formula is highly sensitive to discrete $\mathbb{Z}_2$ symmetry breaking
    or Berezinskii--Kosterlitz--Thouless (BKT) topological transitions~\cite{berezinskii1972destruction,kosterlitz1973ordering,sachdev2011quantum}.
\end{itemize}

\begin{table}[ht]
\centering
\caption{Pairing between LCL formulas and model families used for validation.}
\label{tab:formula_model_pairing}
\resizebox{\linewidth}{!}{

\begin{tabular}{c ccccc >{\centering\arraybackslash}p{1.6cm}
                         >{\centering\arraybackslash}p{1.6cm}
                         >{\centering\arraybackslash}p{1.6cm}}
\toprule
 & \multicolumn{5}{c}{(A) Synthetic channel families} 
 & \multicolumn{3}{c}{(B) Physical spin-chain families} \\
\cmidrule(lr){2-6} \cmidrule(lr){7-9}
Formula 
 & AKLT-like 
 & Cluster-like 
 & Random-gapped 
 & Near-critical 
 & Periodic 
 & TFIM 
 & XXZ 
 & Kitaev \\
\midrule
$\Phi_1$ & \checkmark &  & \checkmark &  &  & \checkmark &  &  \\
$\Phi_2$ & \checkmark &  & \checkmark &  &  & \checkmark &  &  \\
$\Phi_2'$ & \checkmark &  & \checkmark &  &  & \checkmark &  & \checkmark \\
$\Phi_3$ & \checkmark &  & \checkmark &  &  & \checkmark & \checkmark &  \\
$\Phi_4$ &  & \checkmark &  & \checkmark  & \checkmark & \checkmark  &  & \checkmark \\
$\Phi_5$ &  & \checkmark &  &  & \checkmark &  &  &  \\
$\Phi_6$ &  &  &  & \checkmark &  & \checkmark & \checkmark &  \\
\bottomrule
\end{tabular}}
\end{table}

As suggested in Table~\ref{tab:formula_model_pairing},
the validation of LCL specifications demands careful pairing with model families
whose asymptotic transfer-operator structures theoretically guarantee the targeted physical behavior.
\begin{itemize}
    \item \textbf{AKLT and random-gapped models:} The AKLT model is strictly paired with
    formulas testing validity, normalization, and exponential clustering ($\Phi_1, \Phi_2, \Phi_2', \Phi_3$).
    This pairing is physically required because the AKLT state provides a rigorously proven valence-bond ground state
    exhibiting a finite Haldane gap and strict finite correlation lengths~\cite{affleck1988valence}.
    
    \item \textbf{Near-critical and XXZ models:}
    The families are deliberately paired with phase transition indicators ($\Phi_6$)
    and formulas omitting strict clustering bounds.
    As these systems approach a quantum critical point,
    they transition into Tomonaga--Luttinger liquids where exponential clustering breaks down into algebraic decay~\cite{Giamarchi2003}.
    Testing LCL against the models validates the logic's robustness when standard area-law constraints fail.
    
    \item \textbf{Periodic and TFIM models:}
    The Transverse-Field Ising Model (TFIM) and explicitly periodic synthetic channels are paired with formulas evaluating oscillatory constraints ($\Phi_4, \Phi_5$).
    It isolates the peripheral spectrum of the transfer matrix to correctly identify the discrete symmetry
    breaking inherent in the ordered phase of the TFIM~\cite{Perez-Garcia2007}.
\end{itemize}

\section{Proofs Omitted in Main Text}
\subsection{Proof of Lemma~\ref{lem:refine}}
As $\mathcal{H}=\mathcal{H}_1\oplus \mathcal{H}_2$,
we can obtain the two Hermitian projectors $P_1$ and $P_2$,
satisfying $P_1\mathcal{H}=\mathcal{H}_1$, $P_2\mathcal{H}=\mathcal{H}_2$ and $P_1+P_2=I$.
In a basis adapted to this decomposition, each Kraus operator $A_k$ of $\e$ has an upper block-triangular form
\[
A_k=\begin{bmatrix} A_{k,11} & A_{k,12} \\ 0 & A_{k,22}\end{bmatrix}.
\]
Define the block-diagonal part
\[
B_k \coloneqq P_1 A_k P_1 + P_2 A_k P_2
     = \begin{bmatrix} A_{k,11} & 0 \\ 0 & A_{k,22}\end{bmatrix}.
\]
Since $A_{k_1}\cdots A_{k_N}$ remains upper block-triangular and the trace depends only on the diagonal blocks,
we have that for every word $(k_1,\ldots,k_N)$,
\[
\tr(A_{k_1}\cdots A_{k_N})=\tr(B_{k_1}\cdots B_{k_N}).
\]
Consequently, the MPS generated by $\{A_k\}$ coincides with the one generated by $\{B_k\}$,
and in particular their norms satisfy $\tr(M_{\mathcal{E}}^{\,N})=\tr(M_{\widetilde{\mathcal{E}}}^{\,N})$ for all $N$,
where $\widetilde{\mathcal{E}}(X)=\sum_k B_k X B_k^\dagger$.

\subsection{Proof of Lemma~\ref{lem:struct}}
It suffices to consider how to compare the value expression $\Gamma$ on $\ket{\psi_N}$
with any given constant (taken from the endpoints of $\mathcal{I}$),
when the parameter $N$ varies.
For simplification, we assume that the value expression $\Gamma(N)$ is exactly $\braket{\psi_N}{\psi_N}$;
otherwise the employed approach works similarly.
Mathematically, it is exactly the sign determination problem
\[
    \underbrace{c_{1,1}\lambda_{1,1}^N+\cdots+c_{1,d_1}\lambda_{1,d_1}^N}_{\text{largest modulus}} 
    \ + \ \underbrace{c_{2,1}\lambda_{2,1}^N+\cdots+c_{2,d_2}\lambda_{2,d_2}^N}_{\text{second largest modulus}}
    \ + \ \cdots ,
\]
where $c_{i,j}$ are $\mathbb{C}$-coefficients
and $\lambda_{i,j}$ are $\mathbb{C}$-bases from the eigenvalues of $\e$, sorted by their modulus.
Note that the sum of all terms of same modulus is always a real number~\cite{HHH+05}.
We assume w.\,l.\,o.\,g.\@ that the eigenvalues of the largest modulus are unit,
since otherwise they can be normalised to be unit.
Then:
\begin{enumerate}
    \item By Lemma~\ref{lem:cyclic}, all terms of the largest modulus are periodic (with $N$),
    as well as their sum.
    If the sum hits zero periodically,
    we need the terms of the second largest modulus to determine the sign.
    Otherwise the sign determination is plainly completed.
    (There is a computable threshold $N_1$,
    after which the sum of all terms of the largest modulus are exponentially dominating the rest.)
    \item At those MPS in which the sum of all terms of the largest modulus hits zero,
    if the sum of all terms of the second largest modulus is also periodic,
    the case is reduced to the above;
    or if the sum is strictly definite, the case is solved.
    (There is a computable threshold $N_2$,
    after which the sum of all terms of the second largest modulus are exponentially dominating the rest.)
    \item Otherwise, either the sum crosses zero infinitely often,
    which is not exponentially dominating in the ultimate;
    or the sum approaches zero arbitrarily close,
    which is not ensure the computability of the threshold in general,
    as the \emph{effective} ultimate positivity problem is still open~\cite{OuW14a}.
\end{enumerate}
\end{document}